# Fundamentals of magnon-based computing

**Andrii V. Chumak**

*Fachbereich Physik and Landesforschungszentrum OPTIMAS, Technische Universität Kaiserslautern, 67663 Kaiserslautern, Germany*

## 1. Introduction

A spin wave is a collective excitation of the electron spin system in a magnetic solid [1]. Spin-wave characteristics can be varied by a wide range of parameters including the choice of the magnetic material, the shape of the sample as well as the orientation and size of the applied biasing magnetic field [2, 3]. This, in combination with a rich choice of linear and non-linear spin-wave properties [4], renders spin waves excellent objects for the studies of general wave physics. One- and two-dimensional soliton formation [5, 6], non-diffractive spin-wave caustic beams [7-10], wave-front reversals [11, 12], and room temperature Bose–Einstein condensation of magnons [13-16] is just a small selection of examples.

On the other hand, spin waves in the GHz frequency range are of large interest for applications in telecommunication systems and radars. Since the spin-wave wavelengths are orders of magnitude smaller compared to electromagnetic waves of the same frequency, they allow for the design of micro- and nano-sized devices for analog data processing (e.g. filters, delay lines, phase shifters, isolators – see e.g. special issue on Circuits, System and Signal Processing, volume 4, No. 1-2, 1985). Nowadays, spin waves and their quanta, magnons, are attracting much attention also due to another very ambitious perspective. They are being considered as data carriers in novel computing devices instead of electrons in electronics. The main advantages offered by magnons for data processing are [17-21]:

- *Wave-based computing.* If data is carried by a wave rather than by particles such as electrons, the phase of the wave allows for operations with vector variables rather than scalar variables and, thus, provides an additional degree of freedom in data processing [20-29]. This opens access to a decrease of the number of processing elements, a decrease in footprint [20-26], parallel data processing [27], non-Boolean computing algorithms [28, 29], etc.

- *Metal- and insulator-based spintronics.* Magnons transfer spin not only in magnetic metals/semiconductors but also in magnetic dielectrics like the low-damping ferrimagnetic insulator Yttrium-Iron-Garnet (YIG) [30-33]. In YIG, magnons can propagate over centimeter distances, while an electron-carried spin current is limited by the spin diffusing length and does not exceed one micrometer in metals and semiconductors. Moreover, a magnon current does not involve the motion of electrons and thus, it is free of Joule heat dissipation.

- *Wide frequency range from GHz to THz.* The wave frequency limits the maximum clock rate

of a computing device. The magnon spectrum covers the GHz frequency range used nowadays in communication systems [4, 33-35], and it reaches into the very promising THz range [36-38]. For example, the edge of the first magnonic Brillouin zone in YIG lies at about 7 THz [32, 39, 40].

- *Nanosized structural elements.* The minimum sizes of wave-based computing elements are defined by the wavelength λ of the wave used. Among a wide choice of waves in nature, spin waves seem to be one of the most promising candidate since the wavelength of the spin wave is only limited by the lattice constant of a magnetic material and allows for operations with wavelengths down to the few nm regime (probably the first steps in these direction for magnetic insulators were done in [41, 42]). Moreover, the frequency of short-wavelength exchange magnons increases with increasing wavenumber as well as the group velocity [43-45].

- *Pronounced nonlinear phenomena.* In order to process information, non-linear elements are required in order for one signal to be manipulated by another (like semiconductor transistors in electronics). Spin waves have a variety of pronounced nonlinear effects that can be used for the control of one magnon current by another, for suppression or amplification [2-4, 46, 47]. Such magnon-magnon interactions were used for the realization of a magnon transistor and they open access to all-magnon integrated magnon circuits [24].

The field of science that refers to information transport and processing by spin waves is known as *magnonics* [4, 17, 48, 49]. The utilization of magnonic approaches in the field of spintronics, hitherto addressing electron-based spin currents, gave birth to the field of *magnon spintronics* [18, 19, 50]. Magnon spintronics comprises magnon-based elements operating with analog and digital data as well as converters between the magnon subsystem and electron-based spin and charge currents.

The most recent and advanced realizations of spin-wave logic devices including Boolean and Non-Boolean logic gates (e.g. magnonic holographic memory, pattern recognition, prime factorization problem) are addressed in the chapter 19 of Spintronics Handbook: Spin Transport and Magnetism, Second Edition, edited by E. Y. Tsymbal and I. Žutić (CRC Press, Boca Raton, Florida) (volume 3) written by Alexander Khitun and Ilya Krivorotov.

The current chapter addresses a selection of fundamental topics that form the basis of the magnon-based computing and are of primary importance for the further development of this concept. First, the transport of spin-wave-carried information in one and two dimensions that is required for the realization of logic elements and integrated magnon circuits is covered. Second, the convertors between spin waves and electron (charge and spin) currents are discussed. These convertors are necessary for the compatibility of magnonic devices with modern CMOS technology. The chapter starts with basics on spin waves and the related methodology. In addition, the general ideas behind magnon-based computing are presented. The chapter finishes with conclusions and an outlook on the perspective use of spin waves.

## 2. Basics of magnon spintronics

In this section a basic knowledge on spin waves in the most commonly used structure, a spin-wave waveguide in the form of a narrow strip, is given. Formulas that can be used for the calculation of spin-wave dispersions as well for spin-wave lifetime are presented along with the discussions of the main factors that should be taken into account at the micro and nano-scale. In addition, an estimation of the properties of spin waves with wavelength down to a few nanometers is performed in order to give an understanding of the future potential of the field of magnonics. Finally, magnetic materials used in magnonics are discussed along with the methodologies for the fabrication of spin-wave structures as well as for spin-wave excitation and detection.

### 2.1. Spin-wave dispersion relations

The main spin-wave characteristics can be obtained from the analysis of its dispersion relation, i.e. the dependence of the wave frequency $\omega$ on its wavenumber $k$. In the simplest case, there are two main contributors to the spin-wave energy: long-range dipole-dipole and short-range exchange interactions [2, 3]. As a result, the dispersions of spin waves are complex and significantly different from the well-known dispersion of light or sound in uniform media. Moreover, the dispersion relations in in-plane magnetized films are strongly anisotropic due to the dipolar interaction [51]. In most practical situations, spin waves are studied in spatially localized samples such as thin films or strips which are in-plane magnetized by an external magnetic field. The geometry of a spin-wave waveguide, namely its thickness $d$ and its width $w$, is a key parameter defining the spin-wave dispersion along with the effective saturation magnetization $M_S$ of the magnetic material, exchange constant $A_{ex}$, and the applied magnetic field $\mu_0 H$ ($\mu_0 = 1.257 \cdot 10^{-6}$ (T m)/A is the magnetic permeability of vacuum).

In order to calculate spin-wave dispersion relations $\omega(k)$ a theoretical model developed in [52, 53] can be used. This model takes into account both dipolar and exchange interactions as well as spin-pinning conditions at the film surfaces. In order to adopt the model to the case of a spin-wave waveguide of a finite width $w$, a total wavenumber $k_{total} = \sqrt{k_\perp^2 + k_\parallel^2}$, and $\theta_k = \operatorname{atan}\left[ k_\perp / k_\parallel \right]$ should be considered [54, 55]. Here $k_\parallel$ is the spin-wave wavenumber along the waveguide (see inset in Figure 1), $k_{\perp}=n\dfrac{\pi}{w}$ is the perpendicular quantized wavenumber with the number of spin-wave width mode $n$ (please note that in some cases an effective width of the waveguide rather than the real width should be used). In the following $k\equiv k_{\parallel}$ is used underlying that the wave propagating along the waveguide is of importance. The circular frequency of the spin wave can then be presented:

$$\omega(k)=\sqrt{\left(\omega_H+\omega_M\lambda_{ex}\left[k^2+(n\pi/w)^2\right]\right)\left(\omega_H+\omega_M\lambda_{ex}\left[k^2+(n\pi/w)^2\right]+\omega_M F\right)}\,, \qquad (1)$$

where $\omega_H=\gamma\mu_0 H_{eff}$, $\omega_M=\gamma\mu_0 M_S$, $\gamma=1.76\cdot 10^{11}$ rad/(s T) is the gyromagnetic ratio, $\mu_0 H_{eff}$ is the effective internal magnetic field, $M_S$ is the saturation magnetization, $\lambda_{ex}=2A_{ex}/(\mu_0 M_S^2)$ and $A_{ex}$ are the exchange constants. $F$ is given by

$$F=1-g\cos^2(\theta_k-\theta_M)+\dfrac{\omega_M g(1-g)\sin^2(\theta_k-\theta_M)}{\omega_H+\omega_M\lambda_{ex}\left[k^2+(n\pi/w)^2\right]}\,, \qquad (2)$$

where $\theta_k=\mathrm{atan}\left[n\pi/(kw)\right]$ is the angle between the spin-wave wave vector and the long axis of the waveguide – see inset in Figure 1, $\theta_M$ is the angle between the magnetization direction and the long axis of the waveguide (in this model $0\leq\theta_M,\theta_k\leq\pi/2$), $g=1-\left[1-\exp\left(-d\sqrt{k^2+(n\pi/w)^2}\right)\right]/\left(d\sqrt{k^2+(n\pi/w)^2}\right)$. It is important to note that strictly speaking the dispersion relation (1) is valid for the case when $kd<1$ (however, deviations take place in the dipolar-exchange part of the spectrum only, the model works well for the pure exchange waves with large $k$), it takes into account fully unpinned spins at the surfaces of the waveguide, higher-order thickness modes are not considered (in nm-thick samples they are usually few GHz higher in frequency) and magnetic crystallographic anisotropy is omitted (for the account of the anisotropy please see e.g. [56]). Comparison of different approaches to calculate dipolar-exchange spin-wave dispersions can be found in Ref. [57].

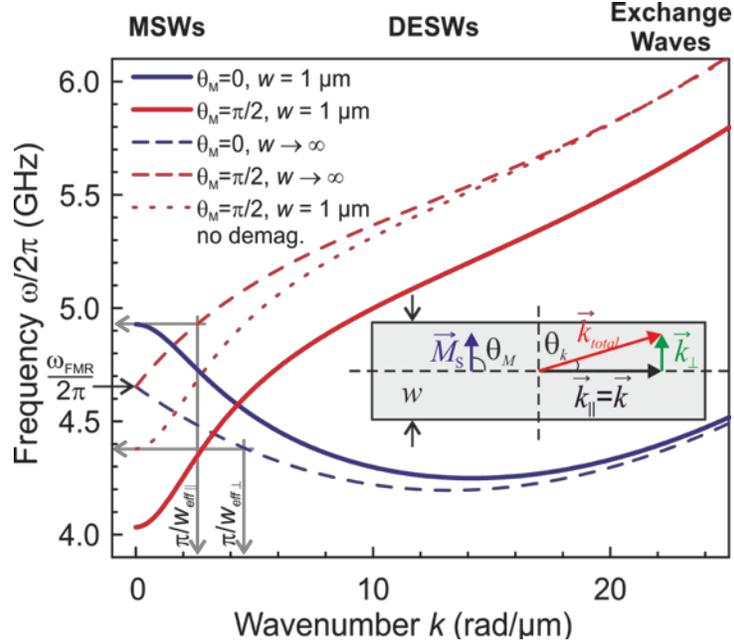

**Figure 1.** Spin-wave dispersion characteristics for an infinite in-plane magnetized YIG film (dashed lines) and spin-wave waveguide $w=1$ µm (solid bold lines) for BVMSW ($\theta_M=0$) and MSSW ($\theta_M=\pi/2$) geometries calculated using Eq. (1). The dotted line shows the spin-wave dispersion in the waveguide in the absence of demagnetization. First width mode $n=1$ is considered, YIG film/waveguide thickness $d=100$ nm, saturation magnetization $M_S=140$ kA/m, exchange constant $A_{ex}=3.5$ pJ/m, a magnetic field of $\mu_0 H_{ext}=100$ mT is applied in-plane. The inset shows the schematics of the spin-wave wavevector components and corresponding angles for the case of MSSW. Dashed lines show long and short axis of a spin-wave waveguide.

Nowadays, spin waves are usually studied in nanometer-thick and micrometer-wide waveguides and, therefore, two additional factors, which define spin-wave properties, should be considered. These are the demagnetization and the spin pinning conditions at the edges of the waveguide. To address these factors, it makes sense to consider waveguides magnetized longitudinally ($\theta_M=0 \Rightarrow \vec{k} \| \vec{M}$) and transversally in-plane ($\theta_M=\pi/2 \Rightarrow \vec{k} \perp \vec{M}$) separately. In the first case, the demagnetization can be ignored since the length of the waveguide is considered to be much larger in comparison to other waveguide dimensions. Therefore, the static internal magnetic field $\mu_0 H_{eff}$ is uniform and is equal to the external applied field $\mu_0 H_{ext}$. The pinning conditions will be mainly defined by the dipolar alternating magnetic fields [58-60] and typically the spins at the edges are partially unpinned. The pinning can be described in terms of the effective width of the waveguide $w_{eff\parallel} \geq w$. If the spins are fully pinned than $w_{eff\parallel} = w$ and if the spins are fully unpinned $w_{eff\parallel} \to \infty$. The effective width (for $\theta_M = 0$) can be found [58]:

$$w_{eff\parallel} = w \frac{D_{Dip}}{D_{Dip} - 2}, \qquad (3)$$

where $D_{Dip} = 2\pi(w/d)/\left[1 + 2\ln(w/d)\right]$.

The case of the transversally magnetized spin-wave waveguide $\theta_M = \pi/2$ is more complex since the external magnetic field has to compete with a shape anisotropy that tends to align the magnetization along the long axis in order to minimize the static stray fields [61, 62]. Therefore, the internal magnetic field is smaller than the external field and is non-uniform showing minima at the edges and a maximum in the center. In the assumption that the magnetization is always aligned along the short axis, the effective internal field can be found as [62]:

$$\mu_0 H_{eff} = \mu_0 H_{ext} - M_s \frac{\mu_0}{\pi}\left[\mathrm{atan}\left(\frac{d}{2z+w}\right) - \mathrm{atan}\left(\frac{d}{2z-w}\right)\right], \qquad (4)$$

where $z$ is the coordinate transverse to the waveguide ($-w/2 \leq z \leq w/2$). Usually, spin waves propagate in the center area of the waveguide with relatively uniform internal magnetic field. (However, waves can also propagate in strongly nonuniform magnetic fields close to the edges of the waveguides that are known as edge modes [41, 54, 55, 63].) In the case of the spin-wave propagation in the center, the internal magnetic field is considered to be uniform with a value $\mu_0 H_{eff}^{max}$ equal to the maximal value given by (4) at $z=0$. The effective width of the waveguide $w_{eff\perp}$, in this case, can be defined in different ways, e.g. as the distance between the points where the effective field is reduced by 10% i.e. to the value $0.9\,\mu_0 H_{eff}^{max}$.

The dispersion relations for an infinitely large plane film (dashed lines) as well as for a spin-wave waveguide (bold solid lines) are shown in Figure 1 for the cases of the spin-wave waveguide magnetized along the long axis $\theta_M = 0$ and transversally $\theta_M = \pi/2$. Only the first width mode $n=1$ is shown for simplicity. The magnetic parameters of the commonly used ferrimagnetic material YIG are considered [41, 42, 64, 65]: Saturation magnetization $M_S = 140\,\mathrm{kA/m}$, exchange constant $A_{ex}$=3.5 pJ/m, the thickness of the waveguide is $d$=100 nm, and its width is $w$=1 µm. The external magnetic in-plane field is $\mu_0 H_{ext}$=100 mT. The spin-wave dispersions comprise three main regions: the region of small wavenumber corresponds to dipolar waves usually termed MagnetoStatic Waves (MSWs) [2, 3], the region of large wavenumbers corresponds to exchange waves, and the region between corresponds to the Dipolar-Exchange Spin Waves (DESWs) – see labels above in Figure 1. The dipolar wave that propagates along the magnetization direction is usually termed as *Backward Volume MagnetoStatic Wave (BVMSW)*, the wave that propagates perpendicularly is named *Magnetostatic Surface Spin Wave (MSSW)* or *Damon Eshbach mode* [2, 3, 51]. (The spin wave propagating in out-of-plane magnetized film is named *Forward Volume MagnetoStatic Waves (FVMSW)* [2, 3, 66] and is beyond the scope of this section.) Please note that with the account of the exchange interaction, the terminology used for the dipolar MSWs is not strictly-speaking applicable. However, even in this case one often uses terms like *BVMSW-* or *MSSW-geometries*. A special frequency, which is indicated in Figure 1a with a black arrow, is the *FerroMagnetic Resonance (FMR)* frequency $\omega_{FMR}/2\pi \approx 4.65$ GHz. The FMR describes the case of a uniform magnetization precession $k_\parallel = k_\perp = 0$. The dispersions for both BVMSW and MSSW modes start from this frequency for the case of a plane film. However, if one considers a waveguide, the frequency of the $k$=0 wave differs from the FMR frequency for both magnetization configurations since in the waveguide there is always $k_\perp > 0$. The frequency of the BVMSW mode $\omega_{\parallel,k=0}/2\pi \approx 4.93$ GHz is defined by the frequency of the first standing mode, which corresponds to an MSSW with $k_\perp = \pi/w_{eff\parallel} = 2.58$ rad/µm and is higher than the FMR frequency shown by the black arrow in Figure 1. Please note that, here, the effective $w_{eff\parallel} \approx 1.22$ µm calculated using Eq. (3) is considered instead of the real waveguide width $w$. It can be seen in the figure that with an increase in $k$, the frequency of the BVMSW mode decreases for both the plane film as well as for the waveguide. This happens due to dipolar interaction and results in a negative group velocity of the dipolar BVMSWs. This negative velocity allows for phenomena like the reverse Doppler shift [67, 68]. With a further increase in $k$, the frequency of the spin wave increases due to the exchange interaction according to the $k^2$ law.

The frequency at the starting point $k$=0 of the MSSW in the waveguide

$\omega_{\perp,k=0}/2\pi=4.03$ GHz lies below the FMR frequency $\omega_{FMR}$. There are two phenomena responsible for such an occurrence. First of all, again, the quantization over the waveguide width plays role. This time one has to consider that the first width mode corresponds to a BVMSW mode with $k_\perp=\pi/w_{eff\perp}=4.62$ rad/µm, where the effective waveguide width $w_{eff\perp}=0.68$ µm calculated according to Eq. (4) and the description below the equation is used. The second mechanism responsible for the further shift of the $\omega_{\perp,k=0}$ frequency down is the demagnetization. According to Eq. (4) the field in the center of the waveguide is $\mu_0 H_{eff}=88.8$ mT and it is smaller than the applied external field $\mu_0 H_{ext}$. In order to demonstrate the contributions of the demagnetization and quantization individually, an "artificial" dispersion in the absence of the demagnetizing field (11.2 mT) is shown in Figure 1a by the dotted line. In Figure 1 the first standing BVMSW mode defines the $\omega_{\perp,k=0,nodemag}=4.38$ GHz frequency. Therefore, the frequency $\omega_{\perp,k=0}$ GHz is well below the corresponding frequency $\omega_{\perp,k=0,nodemag}$ in the absence of the demagnetization. Finally, it can be observed that with an increase in $k$ the spin-wave dispersion characteristics for the plane film and waveguides are getting closer to one another. It is a consequence of the fact that the exchange spin waves are not angle dependent and are less sensitive to the geometry of the waveguide than the dipolar MSWs.

## *2.2. Magnon lifetime and free-path*

If spin waves are to be considered as information carriers in future data processing devices, one can formulate a set of requirements for the spin-wave characteristics [19]. Particularly, the minimum size of a magnonic device should be larger than the wavelengths $\lambda=2\pi/k$ in order to keep access to the usage of phase of the spin waves. The minimization of the delays for data transfer between different magnonic elements requires maximization of the spin-wave group velocity $v_g$. In the simplest case, the clock rate of the devices is limited by the spin-wave frequency $\omega$ and, therefore, the frequency should also be as high as possible. Further, the parasitic loss in the magnetic devices will be inversely proportional to the spin-wave free-path (or propagation length) $l_{free}=v_{gr}\tau$, where $\tau$ is the spin-wave lifetime. Finally, in order to exploit the spin-wave phases for data processing, the ratio of the spin-wave free-path to the wavelength $l_{free}/\lambda$ is of importance.

In general, the minimal wavelength of the spin wave is limited by the lattice constant of a magnetic material and, therefore, magnonics has similar fundamental limitations as electronics. Moreover, the decrease in the wavelength (in the assumption of purely exchange spin waves of wavenumbers staying away from the edge of the Brillouin zone) results in an increase of the frequency $\omega \propto k^2$ of the spin wave and the group velocity $v_{gr} \propto k$. But the spin-wave lifetime $\tau$ is, in the simplest case, inversely proportional to the spin-wave frequency $\tau \propto 1/\omega$ [3] and, thus, the dependence of the spin-wave propagation length on its frequency is $l_{free} \propto 1/\sqrt{\omega}$. However, in the case when both dipolar and exchange interactions are taken into account, the dependence of the free-path on the spin-wave wavenumber $l_{free}(k)$ is not trivial. In order to estimate it, let us consider the same YIG waveguide discussed above magnetized along (this magnetization direction is preferable since it requires minimal external field and ensures uniform internal magnetic field). The lattice constant of YIG is approximately 1.24 nm [30-33, 64] and, in order to stay away from the edge of the first Brillouin zone, the minimal wavelength of 5 nm is considered below. The dispersion relation of the first width mode $n=1$ calculated using (1) is shown in Figure 2a in logarithmic scale (left axis). It can be seen that the spin wave with wavelength of 5 nm has a frequency of about 2 THz in YIG. On the right axis, the corresponding spin-wave wavelengths are shown.

The group velocity of spin waves is defined as $v_{gr} = \partial \omega / \partial k$ and can be found by taking differentiation of Equation (1). The dependence of the velocity on the spin-wave wavenumber is shown in Figure 2b on the left axis. It is clear that in this particular case the group velocity is negative for the BVMSW with small wavenumbers, then it passes through the zero value in the DESW region, and increases monotonically in the exchange region reaching values of around 20 km/s.

The main parameters that define the lifetime $\tau$ of spin wave are the spin-wave frequency $\omega$ and Gilbert damping constant of a magnetic material $\alpha$ [2, 3, 69-71] ($1/\alpha$ defines approximately the number of precession periods before it vanishes). In the simplest case of a circular magnetization precession, the lifetime is simply defined as $\tau=1/(\alpha\omega)$. (Please note that the inhomogeneous broadening of the FMR linewidth $\Delta H_0$ [71] is neglected). However, the magnetization precession in spin-wave waveguides is usually elliptic and the lifetime can be found according to the phenomenological loss theory [3]:

$$\tau=\left(\alpha\omega\frac{\partial\omega}{\partial\omega_H}\right)^{-1}, \qquad (5)$$

with the aforementioned definition $\omega_H = \gamma \mu_0 H_{eff}$. Differentiating Eq. (1) one obtains

$$\tau = \frac{1}{\alpha} \left( \omega_H + \omega_M \lambda_{ex} \left( k^2 + (n\pi/w)^2 \right) + \frac{\omega_M}{2} \left[ 1 - g\cos^2(\theta_k - \theta_M) \right] \right)^{-1} \qquad (6)$$

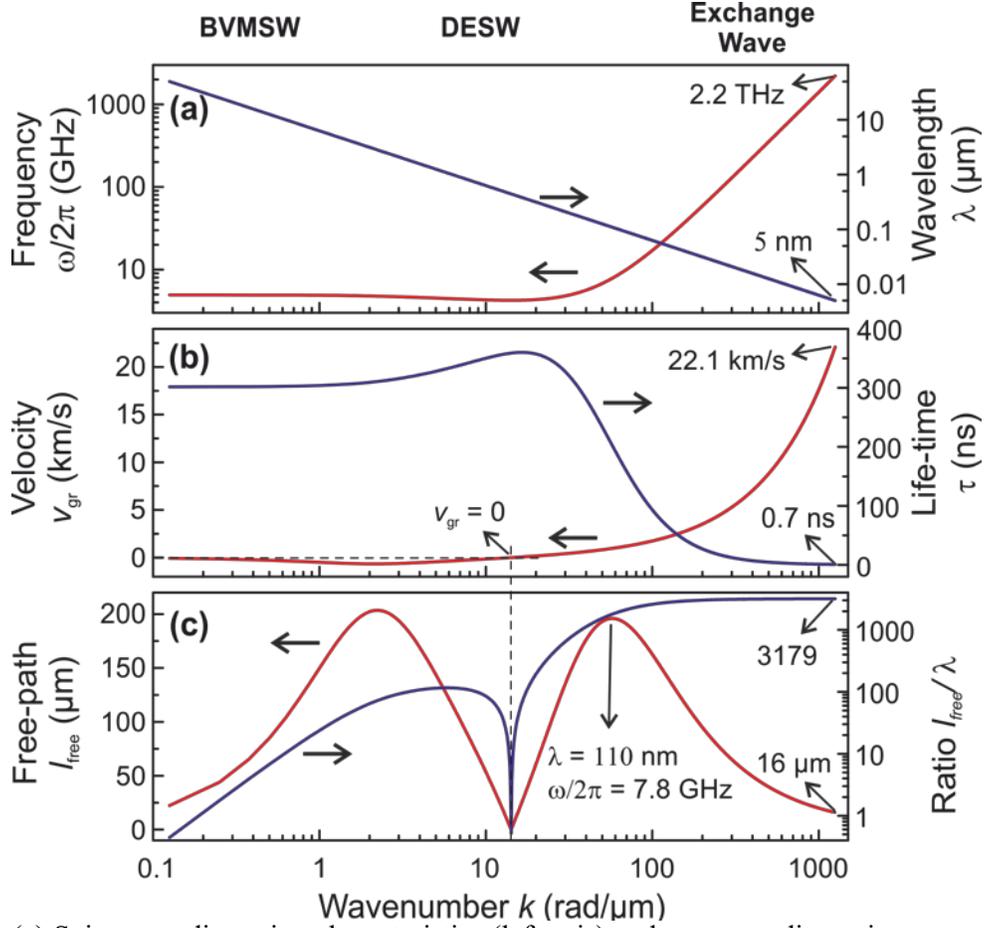

**Figure 2.** (a) Spin-wave dispersion characteristics (left axis) and corresponding spin-wave wavelength (right axis). (b) Spin-wave group velocity (left axis) and lifetime (right axis) as functions of spin-wave wavenumber. (c) Spin-wave free-path (left axis) and the ratio of the free-path to the spin-wave wavelength (right axis) as functions of spin-wave wavenumber. The first width mode $n=1$ is considered in YIG waveguide of width $w=1$ μm and thickness $d=100$ nm magnetized along with $\mu_0 H_{ext}=100$ mT magnetic field. The Gilbert damping parameter of nano-YIG is assumed to be $\alpha=1\cdot 10^{-4}$, which is in agreement with latest reports [64].

The dependence of the lifetime on the wavenumber calculated according to (6) is shown in Figure 2b (right axis) for YIG having a Gilbert damping constant $\alpha = 1 \cdot 10^{-4}$. One sees that in the MSW region the spin-wave lifetime is approximately a few hundred nanoseconds and is practically constant due to the rather flat dispersion and due to the decrease of ellipticity with increasing wavenumber. As opposite, in the exchange region the spin-wave frequency increases rapidly with the wavenumber (Figure 2a) and therefore the lifetime drops down to a value around 1 ns.

The spin-wave free path (distance which spin wave propagates before its amplitude decreases to its $1/e$ value) $l_{free} = v_{gr} \tau$ is shown in Figure 2c on the left axis. It can be clearly observed that the free-path of the long-wavelength BVMSW is rather large and is close to few hundreds micrometers. Moreover, $l_{free}$ is almost proportional to the film thickness $t$ for MSWs and, therefore, this value is much larger in the µm-thick YIG samples that were intensively studied in the recent decades [4]. However, the miniaturization of magnonic elements requires a decrease in the thickness of the structures to the nm scale and the propagation distance of MSWs drastically reduces. Therefore, exchange rather than dipolar waves are primarily of interest for future investigations [43-45] (although nowadays experimental magnonics mainly operates with dipolar MSWs due to relative simplicity in the methodology [4, 19, 54, 72]). It is seen in Figure 2c, that the free-path of DESW is close to zero due to its zero group velocity [13, 73, 74] (please note that it is not the case for MSSWs – see the dispersion slope in Figure 1). With a further increase in $k$, the spin-wave group velocity increases along with a decrease in the lifetime. Therefore, there is a maximum of the free-path for waves featuring of wavelengths of about 100 nm. A further increase in $k$ results in the decrease in $l_{free}$. Nevertheless, the decrease in the wavelength assumes that long free-paths are not always necessary. The ratio $l_{free}/\lambda$, which shows how many wavelengths (i.e. how many unit elements) a wave propagates before it relaxes, is of importance. It can be clearly seen in the Figure 2c (right axis) that the decrease in the wavelength of the exchange waves results in the increase of the ratio $l_{free}/\lambda$. This ratio reaches values above 3000 for waves of nanometer wavelength. This estimation is very encouraging for the further development of the field of magnonics since it shows that potentially only around $1-\exp(-2\lambda/l_{free}) \approx 6 \cdot 10^{-4}$ of the spin-wave energy will be lost for data transport in a unit element of size $\lambda$.

## *2.3. Materials for magnonics and methodology*

As was discussed above, spin waves are usually studied in thin magnetic films or waveguides in the form of narrow strips. The choice of the material plays a crucial role in fundamental as well as in applied magnonics. The main requirements for the magnetic materials are: (*i*) small Gilbert damping parameter in order to ensure long spin-wave lifetimes; (*ii*) large saturation magnetization for high spin-wave frequencies and velocities in the dipolar region; (*iii*) high Curie temperatures to provide high thermal stability; and (*iv*) simplicity in the fabrication of magnetic films and in the patterning processes [75]. The most commonly used materials for magnonics as well as those with a high potential for magnonic applications are presented in Table 1 (adopted from [75]) together with some selected parameters and estimated spin-wave characteristics: lifetime $\tau$, velocity $v_{gr}$, free-path $l_{free}$, and the ratio $l_{free}/\lambda$. Only the dipolar MSW is considered since for experimental investigations nowadays these are the waves that are mainly used (please note that the values in Ref. [75] differ a bit from the values here since in Table 1 the exchange interaction and the ellipticity of magnetization precession is taken into account). It has to be noted that here MSSW propagating in a plane film perpendicularly to the magnetization orientation rather than BVMSW is considered since it features higher values of the group velocity.

The first material in the table is monocrystalline Yttrium Iron Garnet $Y_3Fe_5O_{12}$ (YIG) films grown by high-temperature liquid phase epitaxy (LPE) on Gadolinium Gallium Garnet (GGG) substrates [30-33, 64, 65, 76, 77]. This ferrimagnet was first synthesized in 1956 by Bertaut and Forrat [76] and has the smallest known magnetic loss that results in the lifetime of spin waves being of some hundreds of nanoseconds and, therefore, finds widespread use in academic research [4, 19, 33, 75]. Many of the experimental results presented in this chapter were obtained using LPE YIG. The small magnetic loss is due to the fact that YIG is a magnetic dielectric (ferrite) with very little spin-orbit interaction and, consequently, with small magnon-phonon coupling [2, 32]. Moreover, high quality LPE single-crystal YIG films ensure a small number of inhomogeneities and, thus, suppressed two-magnon scatterings [2, 91]. However, the thickness of these films, which is in the micrometer range, does not allow for the fabrication of YIG structures of nanometer sizes. Therefore, the fabrication of nanostructures became possible only within the last few years with the development of technologies for the growth of high-quality nm-thick YIG films, see second column in Table 1, by means of e.g. pulsed-laser deposition (PLD) [42, 78-81], sputtering [82], or via modification of the LPE growth technology [41, 64]. Although the quality of these films is still worse when compared to micrometer-thick LPE YIG films, it is already good enough to satisfy the majority of requirements of magnonic applications [19].

|  | μm-thick LPE Yttrium Iron Garnet (YIG) | nm-thick Yttrium Iron Garnet (YIG) | Permalloy (Py) | CoFeB | Heusler CMFS compound |
|---|---|---|---|---|---|
| Chemical composition | $Y_3Fe_5O_{12}$ | $Y_3Fe_5O_{12}$ | $Ni_{81}Fe_{19}$ | $Co_{40}Fe_{40}B_{20}$ | $Co_2Mn_{0.6}Fe_{0.4}Si$ |
| Structure | single-crystal | single-crystal | poly-crystal | amorphous | single-crystal |
| Gilbert damping $\alpha$ | $5 \cdot 10^{-5}$ | $2 \cdot 10^{-4}$ | $7 \cdot 10^{-3}$ | $4 \cdot 10^{-3}$ | $3 \cdot 10^{-3}$ |
| Sat. magnetization $M_0$, kA/m | 140 | 140 | 800 | 1250 | 1000 |
| Exchange constant $A$, pJ/m | 3.6 | 3.6 | 16 | 15 | 13 |
| Curie temperature $T_C$, K | 560 | 560 | 550-870 | 1000 | > 985 |
| Typical film thickness $t$ | 1-20 μm | 5-100 nm | 5-100 nm | 5-100 nm | 5-100 nm |
|  | The following parameters are calculated for dipolar MSSW modes, film magnetized in-plane by the field of 100 mT, for the spin-wave wavenumber $k = 0.1/t$: ||||| 
| Lifetime for dipolar surface wave $\tau$, | 604.9 ns (@ 4.77 GHz) | 150.2 ns (@ 4.8 GHz) | 1.3 ns (@ 11.1 GHz) | 1.6 ns (@ 14.9 GHz) | 2.6 ns (@ 12.8 GHz) |
| Velocity $v_{gr}$ | 33.7 km/s (@ $t = 5$ μm) | 0.23 km/s (@ $t = 20$ nm) | 2.0 km/s (@ $t = 20$ nm) | 3.5 km/s (@ $t = 20$ nm) | 2.6 km/s (@ $t = 20$ nm) |
| Freepath $l$ | 20.4 mm | 35.1 μm | 2.7 μm | 5.7 μm | 6.9 μm |
| Ratio $l/\lambda$, | 64.9 | 27.9 | 2.1 | 4.5 | 5.5 |
| References | [4, 30-33, 64, 65 76, 77] | [41, 42, 64, 78-82] | [54, 71, 72, 83] | [84-86] | [87-90] |

**Table 1.** Shows a selection of magnetic materials for magnonic applications, their main parameters, and estimated spin-wave characteristics. The characteristics are calculated using the dipolar approximation for infinite films magnetized in-plane with a 100 mT magnetic field. The lifetime is estimated with ellipticity being taken into account according to Eq. (6) and for the case of non-uniform FMR linewidth widening to be zero.

The second most commonly used material in magnonics is Permalloy that is a polycrystalline alloy of 80% Ni – 20% Fe (see Table 1). This is a soft magnetic material with low coercivity and anisotropies. One of the major advantages of this material is that it has a fairly low spin-wave damping value considering it is a metal, and it can be easily deposited and nano-structured. Therefore, Permalloy was intensively used for the investigation of spin-wave physics in micro-structures (see reviews [54, 72]). Nowadays, a large quantity of attention from the community is also focused on CoFeB and half-metallic Heusler compounds. These materials possess smaller Gilbert damping parameters and larger values of the saturation magnetization, and, are therefore, even more suitable for the purposes of magnonics. For example, it was demonstrated that the spin-wave mean free path in Heusler compounds can reach 16.7 μm [90].

The fabrication of high-quality spin-wave waveguides in the form of magnetic strips is also one of the primary tasks in the field of magnonics. The most commonly used technique for the fabrication of micrometer-thick YIG waveguides is a dicing saw [92] since the width of the waveguide is usually larger than 1 mm. A commonly used method for the patterning of such YIG films is Photolithography with subsequent wet etching by means of hot Orthophosphoric acid [7, 93]. Different techniques are used to pattern nanometer-thick YIG films: E-beam Lithography with subsequent $Ar^+$ dry etching have shown good results [41, 42]. Focused Ion Beam (FIB) milling has also recently shown very promising results and allows for the fabrication of YIG structures with lateral sizes below 100 nm (not published). The same techniques can also be used for the patterning of metallic magnetic films. Frequently an alternative approach is used; the magnetic material is deposited on a sample covered with a resist mask produced via photo- or electron beam lithography followed by a standard lift-off process. Antennas and the required contact pads are deposited afterwards in subsequent lithography, electron beam evaporation (or sputtering), and lift off processes.

Modern magnonics consists of a wide range of instrumentation for the excitation and detection of magnons. The main requirements for magnon detection techniques could be defined as sensitivity, the range of detectable wavelengths and frequencies, as well as frequency, spatial, and temporal resolution. For the spin-wave excitation techniques, the efficiency of excitation, its coherency as well as the wavenumber range is of primary importance. A broad scope of techniques intensively used nowadays in magnonics as well as techniques showing much potential are listed in [75]. Among others, one can define three main categories of these techniques: microwave approaches, optical technologies, and spintronics approaches. The last one is in the heart of magnon spintronics and is discussed later in more details. One can attribute to the microwave approaches the following techniques: conventional microstrip (or CoPlanar Waveguide (CPW) or meander-type) antennas based techniques for spin wave excitation and detection [4, 94-98], contactless antenna based approaches to excite coherent spin waves [44, 99, 100], FMR spectroscopy [65, 71, 83, 101], parametric pumping technique that is usually used for spin-wave amplification [4, 13-16, 43, 102, 103], Pulsed Inductive Microwave Magnetometer technique (PIMM) [71, 104], and Inductive Magnetic Probe (IMP) technique for the detection [105]. The most commonly used optical techniques are Brillouin Light Scattering (BLS) spectroscopy for the detection of spin waves [54, 72, 106, 107], thermal and non-thermal excitation of spin waves by femtosecond pulsed laser techniques [49, 108-110], and Magneto Optical Kerr Effect (MOKE) spectroscopy for spin-wave detection [99, 100, 110]. With spintronics approaches one can associate Spin Transfer Torque (STT) based techniques used for amplification and generation of spin waves [19, 111-120], Spin Pumping (SP) based approaches for spin-wave detection [19, 43, 121-125], and Spin-Polarized Electron Energy Loss Spectroscopy (SPEELSC) [126, 127]. Besides the mentioned techniques, there is a set of other promising approaches: Magneto-Electric (ME) cells [20-23, 128, 129], Magnetic Resonance Force Microscopy (MRFM) [130], detection of magnon-induced heat [131], Nuclear Resonant Scattering of synchrotron radiation (NRS) [132], X-ray detected FMR (XFMR) [133, 134], as well as electron-magnon scattering approaches [135, 136].

The most commonly used technique for spin-wave excitation is inductive excitation with a microwave current sent through a strip-line or CPW antenna. In order to understand the spin-wave signal excitation mechanism, it is useful to consider the waveguide as a reservoir of quasi-classical spins. When the waveguide is magnetically saturated, the mean precessional axis of all the spins is parallel to the bias field. The application of a microwave signal to the strip-line antenna generates an alternating Oersted magnetic field around it. The components of this field which are perpendicular to the bias direction, create an alternative torque on the magnetization that results in an increase in the precessional amplitude. The spins precessing under the antenna interact with their nearest neighbors and, if the correct conditions for field and frequency are satisfied, spin-wave propagation is supported. After the propagation, the spin waves might be detected by an identical output microstrip (or CPW or meander-like) antenna [4, 95-98]. The mechanism of spin-wave detection is, by symmetry the inverse of the excitation process.

One of the most powerful techniques in magnonics nowadays is Brillouin Light Scattering (BLS) spectroscopy [4, 54, 72, 106, 107]. The physical basis of BLS spectroscopy is the inelastic scattering of photons by magnons. Scattered light from a probe beam, incident on the sample is analyzed and allows the frequencies and wavenumbers of the scattering magnons to be determined, where the scattered photon intensity is proportional to the spin-wave intensity. The technique is generally used in conjunction with a microwave excitation scheme and, over the last decade, has undergone extensive improvements. BLS spectroscopy now achieves a spatial resolution of 250 nm, and time-, phase-, and wavenumber resolved BLS spectroscopy have been realized.

## *2.4 Basic ideas of magnon-based computing*

One of the main strengths of magnonics lies in the benefits provided by the wave nature of magnons for data processing and computation. In the past, the application area of spin waves was mostly related to analogue signal processing in the microwave frequency range. For this applications microwave filters, delay lines, phase conjugators, power limiters, and amplifiers are just a few examples [4, 33-35]. Nowadays, new technologies, allowing, e.g., for the fabrication of nanometer-sized structures or for operation in the THz frequency range, in combination with novel physical phenomena, provide a new momentum to the field and make the advantages discussed earlier accessible for both analogue and digital data signal processing. Magnons also possess the potential to be used in the implementation of alternate computing concepts such as non-Boolean computing [137, 138], reversible logic [28, 139], artificial neural networks [29], and, more general, the projecting of optical computing concepts [140] onto the nanometer scale. These directions are still in the beginning of their development. The basic ideas behind the standard Boolean logic operations with digital binary data, which are currently the subject of intensive theoretical and experimental studies [19], are addressed here. The more advanced developments in magnon-based computing are presented in chapter 19 "Spin Wave Logic Devices" (volume 3) of Spintronics Handbook: Spin Transport and Magnetism, Second Edition, edited by E. Y. Tsymbal and I. Žutić (CRC Press, Boca Raton, Florida).

The idea of coding binary data into spin-wave amplitude was first stated by Hertel et al. in 2004 [141]. Micromagnetic simulations revealed that magnetostatic spin waves change their phase as they pass through domain walls. It was suggested to split spin waves in different branches of a ring (spin-wave interferometer). After being merged in the output, their interference depends on the presence of domain walls in the branches. In such a way, a controlled manipulation of phases of spin waves was proposed to be utilized for the realization of spin-wave logical operations. The first experimental studies of spin-wave logic were reported by Kostylev at al. in Ref. [142]. It was proposed that a Mach-Zehnder spin-wave interferometer equipped with current-controlled phase shifters embedded in the interferometer arms can be used to construct logic gates. Following this idea, Schneider et al. realized a proof-of-principle XNOR logic gate shown in Figure 3a [143]. The currents applied to Input 1 and Input 2 represent logic inputs: The current-on state, which results in π-phase shift of the spin wave in the interferometer arm, corresponds to logic "1"; the current-off state corresponds to logic "0". The logic output is defined by the interference of microwave currents induced by the spin waves in the output microstrip antennas: A low amplitude (destructive interference) corresponds to logic "0", and a high amplitude (constructive interference) defines logic "1". The change in the magnitude of the output pulsed signal for different combinations of the logic inputs (see Figure 3a) corresponds to the XNOR logic functionality. It was also shown that an electric current can create a magnetic barrier reducing or even stopping the spin-wave transmission. Using this effect also a universal NAND logic gate was demonstrated [143]. Lee at al. [26] has proposed an alternative design of a nanometer-sized logic gate – see Figure 3b. In this device, the spin-wave phases are controlled by way of an electric current flowing through a vertical wire placed between the arms of the interferometer. This current defines the logic input, while the amplitude of the output spin wave after interference defines the logic output. The ability to create NOT, NOR and NAND logic gates was demonstrated using numerical simulations [26].

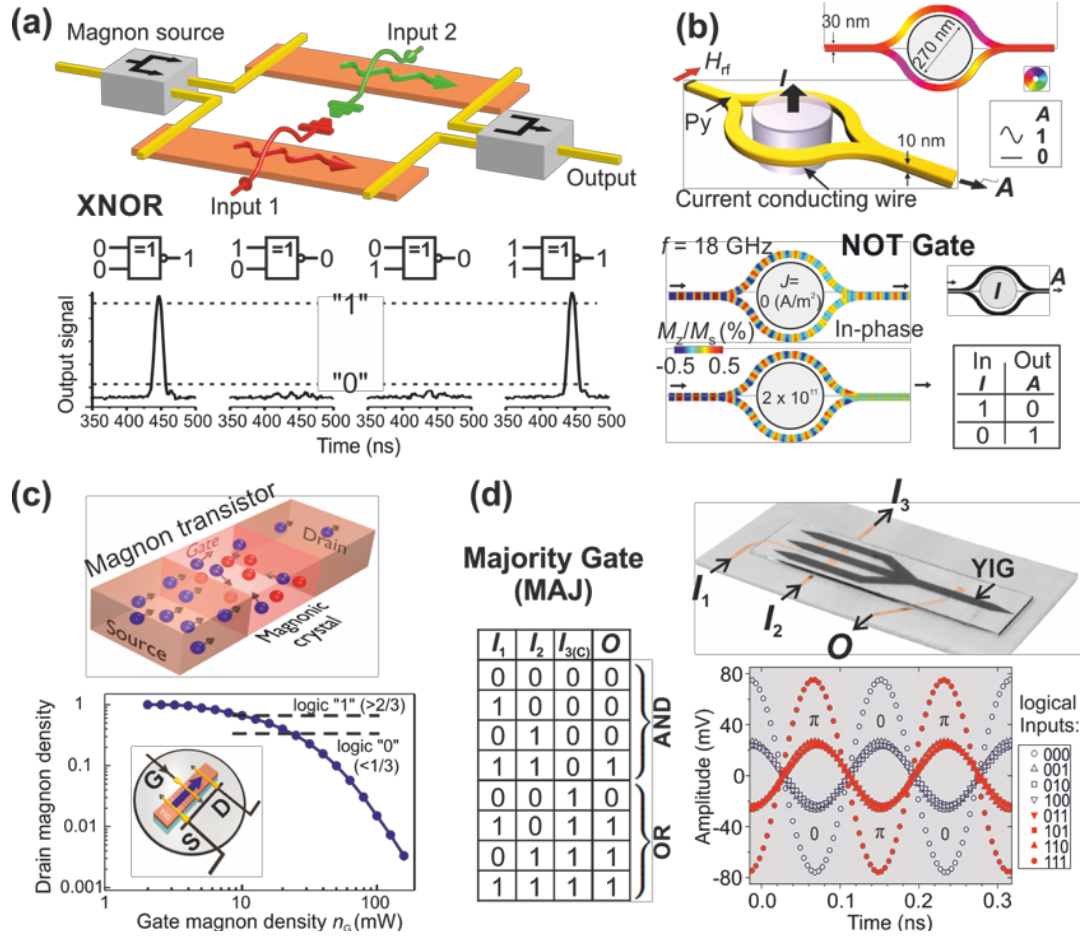

**Figure 3.** (a) Spin-wave XNOR logic gate [18, 143]. The gate is based on a Mach–Zehnder interferometer with electric current-controlled phase shifters embedded in the two magnon conduits. The bottom panel shows the output pulsed spin-wave signals measured for different combinations of the input DC signals applied to the phase shifters. (b) Nanosized Mach–Zehnder spin-wave interferometer designed in a form of a bifurcated Py conduit girdling a vertical conducting wire of 270 nm in diameter (adapted from [26]). Numerical simulation of a NOT logic operation is shown below. (c) Schematic of the operational principle of a magnon transistor [24]: The source-to-drain magnon current (shown with blue spheres) is nonlinearly scattered by gate magnons (red spheres) injected into the gate region. In the bottom panel, the measured drain magnon density is presented as a function of the gate magnon density. The horizontal dashed lines define the drain density levels corresponding to a logic "1" and a logic "0". (d) Left panel: Truth table of the majority operation. Upper panel: photograph of the majority gate prototype under test in Ref. [25]. Spin waves are excited with microstrip antennas ($I_1$, $I_2$, and $I_3$) in the input YIG waveguides (width 1.5 mm) and propagate through the spin-wave combiner into the output waveguide for the detection with the output antenna (O). Lower panel: Output signal observed with an oscilloscope for different phases of the input signals. The dependence of the output phase on the majority of the input phases is clearly visible.

The drawback of the discussed logic gates is that it is impossible to combine two logic gates without additional magnon-to-current and current-to-magnon converters. This fact stimulates the search for a means to control a magnon current by another magnon current. It has been demonstrated that such control is possible due to nonlinear magnon-magnon scattering, and a *magnon transistor* was realized [24] – see Figure 3c. In this three-terminal device, the density of the magnon current flowing from the Source to the Drain (see blue spheres in the Figure) is controlled by the magnons injected into the Gate of the transistor (red spheres). A magnonic crystal (discussed in the next section) in the form of an array of surface grooves [93] is used to increase the density of the gate magnons and, consequently, to enhance the efficiency of the nonlinear four-magnon scattering process used to suppress the Source-to-Drain magnon current. It was shown that the Source-to-Drain current can be decreased by up to three orders of magnitude (see bottom panel in Figure 3c). The potential for the miniaturization of this transistor, for the realization of an integrated magnonic circuit (using the example of an XOR logic gate) as well as for the increase of its operation speed and decrease in energy consumption are discussed in Ref. [24]. Although the operational characteristics of the presented insulator-based transistor in its proof-of-principle form do not overcome those of semiconductor devices, the presented transistor might play an important role for future *all-magnon technology* in which information will be carried and processed solely by magnons [24]. The main advantage of the all-magnon approach is that it does not require convertors between magnons and charge currents in each unit (the converters are needed only in the beginning to code the electric signal into the magnons in a magnonic chip and at the end to read the data out after the processing). The naturally-strong nonlinearity of spin waves is assumed to be used to perform operations with data.

In the approaches discussed above, logic data was coded into spin-wave amplitude (a certain spin-wave amplitude defines logic "1", zero amplitude correspond to logic "0"). Alternatively, Khitun et al. proposed [21, 128, 144] to use the spin-wave *phase* for digitizing information instead of the *amplitude*. A wave with some chosen phase $\phi_0$ corresponds to a logic "0" while a logic "1" is represented by a wave with phase $\phi_0 + \pi$. Such an approach allows for a trivial embedding of a NOT logic element (which requires two transistors in modern CMOS technology) in magnonic circuits by simply changing the position of a read-out device by a $\lambda/2$ distance. Moreover, it opens up access to the realization of a *majority gate* in the form of a multi-input spin-wave combiner [21]. The spin-wave majority gate consists of three input waveguides (generally speaking, of any odd number of inputs larger than one) in which spin waves are excited. A spin-wave combiner, which merges the different input waveguides, and an output waveguide in which a spin wave propagates with the same phase as the majority of the input waves. Thus, the majority logic function can be realized due to a simple interference between the three input waves – see the truth table in the left panel of Figure 3d. The majority gate can perform not only the majority operation but also AND and OR operations (as well as NAND and NOR operations if to use a half-wavelength long inverter), if one of its inputs is used as a control input [145, 146]. The large potential of the majority gate is also underlined by the fact that a full adder (used in electronics to sum up three bits) can be constructed using only 3 majority gates while a total of a few tens of transistors are used nowadays in CMOS. This will allow for a drastic decrease in the footprint of magnonic devices when compared to electronics. One more advantage of the majority gate is that it might operate with spin waves of different wavelengths simultaneously, paving the way towards single chip parallel computing [27]. This approach requires the splitting of the signals of the spin waves that have different wavelengths and, therefore, the usage of magnonic crystals or directional coupler (discussed later) is needed.

The first fully-functioning design of the majority gate employing a spin-wave combiner was demonstrated using micromagnetic simulations by Klingler et al. in Ref. [145]. One of the main problems of a spin-wave majority gate based on the spin-wave combiner is the coexistence of different spin-wave modes with different wavelengths at the same frequency. The non-uniformities of the combiner area, through which spin waves propagate, usually act as re-emitters of new spin waves of the same frequency but having different wavenumbers. By choosing a smaller width for the output waveguide (1 µm in this case as opposite to 2 µm-wide input waveguides), it was possible to select the first width mode from the combiner and to ensure the readability of the output signal. The functionality of the majority gate was proven by showing that all excitation combinations with a majority phase of 0 are in phase and, simultaneously, in anti-phase to all combinations with a majority phase of $\pi$. However, the output signal in [145] was still influenced by exchange spin waves of the same frequency but with shorter wavelengths. To overcome this limitation, isotropic forward volume magnetostatic spin waves (FVMSWs) in an out-of-plane magnetized spin-wave majority gate were used in Ref. [146]. A high spin-wave transmission through the new asymmetric design of the majority gate of up to 64%, which is about three times larger than for the in-plane magnetized gate, was achieved. The phases of output spin wave clearly satisfied the majority function (see the truth table in Figure 3d), proving the functionality of the gate.

Recently, an experimental prototype of a majority gate utilizing macroscopic YIG structure was shown by Fischer et al. [25]. The respective device comprises three input lines, as well as one output line and has been structured from a YIG film of 5.4 µm thickness by means of photolithography and wet chemical etching – see top panel in Figure 3d. In this geometry, inductively excited spin waves propagate coherently along the three input waveguide and eventually superimpose when they leave the spin-wave combiner, at which point the input lines merge. The resultant spin wave induces an electrical signal in a copper stripline at the output which is directly mapped by a fast oscilloscope [25]. The logical information encoded in the phase of the spin waves is controlled by upstream adjustable phase shifters. The lower panel of Figure 3d shows the measured spin-wave amplitudes for all possible combinations of logic input states. The phase of the output spin wave corresponds to the majority phase of the input lines, according to the truth table of a majority gate. Nevertheless, one can clearly see that although the phase of the output signal is defined according to the majority logic function, the amplitude of the output also depends on the combination of the phases of the input spin-wave signals (it is three times larger for the case 0-0-0, when all waves are in phase, in comparison to, e.g., the 0-0-1 case). Therefore, if to follow the all-magnon computing approach, a nonlinear amplitude normalizer is required in order to be able to send the output spin wave from one majority gate as an input wave into the next majority gate in an integrated magnonic circuit. Another way to go was originally proposed by Khitun [21] and assumes that the data is converted from spin waves into an electric signal using magneto-electric cells after each majority gate and is coded back into spin waves in the next element (see the section of magneto-electric cells later). In this case, the nonlinearity required for all-magnon computing is replaced by the conversion itself. However, many state-of-the-art approaches to convert AC spin waves into DC voltages, such as the Inverse Spin Hall Effect or the Tunneling Magnetoresistance Effect, are not sensitive to the spin-wave phase. Hence, the electrical readout of the phase information is a challenge. In Ref. [147] the authors have demonstrated the conversion of the spin-wave phase into a spin-wave intensity by local non-adiabatic parallel pumping (see discussion above). The first step for the practical application of phase-to-intensity conversion was realized experimentally for spin waves in a microstructured magnonic waveguide made from Permalloy [147]. It was also shown how phase-to-intensity conversion can be used to extract the majority information from an all-magnonic majority gate.

## 3. Guiding of spin waves in one and two dimensions

The main problems faced as it rates to the transfer of data between information processing elements in two-dimensional spin-wave circuits are discussed in this section. A selection of perspective solutions for the guiding as well as for the processing of data is presented.

### 3.1. Magnonic crystals

Magnonic crystals are artificial magnetic media with periodic variation of their magnetic properties in space. Bragg scattering affects the spin-wave spectrum of such a structure. This leads to the formation of band gaps – regions of the spin-wave spectrum in which spin-wave propagation is prohibited (see right panels in Figures 4a and 4b). Consequently, areas between band gaps allow for selective spin-wave propagation [4, 49, 75, 106, 148]), while the pronounced changes of the spin-wave dispersion near the band-gap edges opens up access to formation of band-gap solitons [149], the deceleration of spin waves and the appearance of confined spin-wave modes [150]. In spite of the fact that the term "magnonic crystal" is relatively new (it was first introduced by Nikitov et al. in 2001 [151]) this field goes back to the studies of spin-wave propagation in periodical structures that had already been initiated by Sykes, Adam and Collins in 1976 [152]. However, the early magnonic crystal research activities were mainly devoted to the fabrication of microwave filters and resonators (see reviews [153-155]).

Nowadays, when the coupling of spin-wave modes and demagnetizing effects cannot be neglected on the nano- and micro-scales, many of the magnonic crystal studies are focused on the understanding of the spin-wave physics in the crystals. Besides, recent experimental studies of nonlinear magnonic crystals (see for example [4, 149, 156, 157]) as well as the development of magnonic crystal theories (the model of three-dimensional magnonic crystals [158], for example) have brought sufficient progress to the general wave physics. Thus, the field of magnonic crystal is growing rapidly and magnonic crystals have already been given a partial overview in a set of review papers. Such papers include spin-wave dynamics in periodic structures for microwave applications [33, 153-155], magnonics crystals for data processing in general [4, 19], Brillouin light scattering studies of planar metallic magnonic crystals [106], micromagnetic computer simulations of width-modulated waveguides [159], photo-magnonic aspects of the antidot lattices studies [49], theoretical studies of one-dimensional monomode waveguides [160], and reconfigurable magnonic crystals [148].

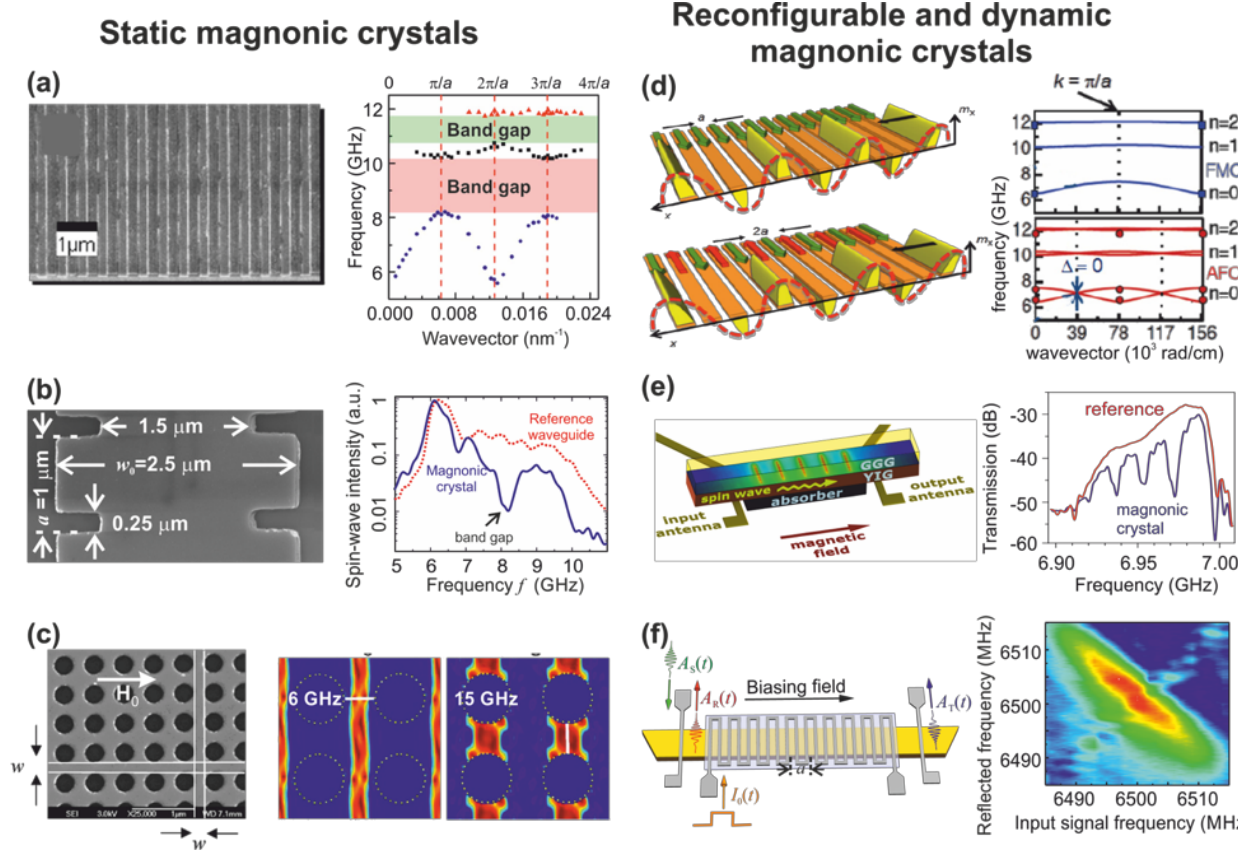

**Figure 4.** Realizations of magnonic crystals. (a) One-dimensional magnonic crystal in the form of an array of alternating Py and Co nanostrips and the structure of magnon band gaps measured by BLS spectroscopy [161]. (b) Magnonic crystal designed as a Py conduit with a periodically varying width, and spin-wave spectra of this crystal and of a uniform reference conduit [162]. (c) Scanning electron microscopy image of a two-dimensional magnonic crystal in the form of an anti-dot array in a Py film and the numerically simulated spatial distribution of spin-wave amplitudes for different magnon modes [164]. (d) Reconfigurable magnonic crystal in the form of an array of bi-stable magnetic nanowires and spin-wave spectra obtained for ferromagnetic and antiferromagnetic nanowires orders [148, 171]. (e) Schematic of the reconfigurable magnonic crystal consisting of a GGG/YIG/absorber multilayer system (temperature is coded in color) and measured spin-wave transmission characteristics in the thermal landscape and reference data without the projected pattern [167]. (f) Schematics of a dynamic magnonic crystal in the form of current-carrying meander structure positioned close to the surface of a YIG conduit [165, 174]. A two-dimensional map of reflected signal spectra as a function of the incident signal frequency demonstrating frequency inversion is shown on the right.

The wide variety of parameters, which define the characteristics of spin waves in a magnetic film, results in a variety of possible designs of magnonic crystals. For example, magnonic crystals can be constructed using two different magnetic materials (bi-component magnonic crystals [161]) – see Figure 4a. Other examples are spin-wave conduits with a periodic variations of their width (Figure 4b) [159, 162, 163], periodic dot or anti-dot lattices (Figure 4c) [49, 164], arrays of interacting magnetic strips (Figure 4d) [106, 148], periodic variations of biasing magnetic field (Figure 4e) [165, 166], temperature (Figure 4f) [167], thickness [4, 152, 168], saturation magnetization [169], mechanical stress [68, 170], etc. When the classifications of magnonic crystals is spoken of, one could define at least two approaches. First of all, based on their dimensions, magnonic crystals can be devided into *one-, two- and three-dimensional magnonic crystals* types. Another way to systematize the crystals is based on the possibility to vary their parameters with time. The most common are *static magnonic crystals* properties of which are defined by the design of the structure and cannot be changed after their fabrication. *Reconfigurable magnonic crystals*, whose properties can be changed on demand [148, 167, 171, 172], attract special attention since they allow for the tuning of the functionality of a magnetic element. The same element can be used in applications as a magnon conduit, a logic gate, or a data buffering element. An example of such a structure is a magnonic crystal in the form of an array of magnetic strips magnetized parallel or anti-parallel to each other [171] – see Figure 4d. Another way for the creation of an arbitrary 2D magnetization pattern in a magnetic film is based on laser-induced heating and was also used to demonstrate a reconfigurable magnonic crystal [167] – see Figure 4e. Furthermore, changes in the properties of a magnonic crystal give access to novel physics, if these changes occur on a timescale shorter than the spin-wave propagation time across the crystal. Such magnonic crystals are termed *dynamic magnonics crystals* [165, 166, 173]. The first dynamic magnonic crystal was realized in the form of a YIG conduit placed in a time-dependent spatially periodic magnetic field. The field was induced by an electric current sent through a meander-type conducting structure placed close to the surface of the conduit - see Figure 4f and could be changed on a timescale below 10ns [165]. It has been shown that this dynamic magnonic crystal can perform a set of spectral transformations such as frequency inversion and time reversal [174]. Finally, another type of dynamic magnonic crystals, in which a "static" magnonic crystal is moving with respect to the spin waves of a certain velocity, is termed *travelling magnonic crystals* [68, 170]. In this case, one should consider a Doppler effect, in addition to the Bragg scattering. As a result, the band gaps of a travelling magnonic crystal are shifted in frequencies and the velocity of the moving grating defines the frequency shift. One of the realization of such crystal is based on the utilization of Surface Acoustic Waves (SAWs) that act as a moving periodic scatterer for spin waves in a YIG film [68, 170].

For magnonics applications, magnonic crystals constitute one of the key elements since they open up access to novel multi-functional magnonic devices [75]. These devices can be used as spin-wave conduits and filters (in fact, any magnonic crystal can serve as a conduit or a filter), sensors, delay lines and phase shifters, components of auto-oscillators, frequency and time inverters, data-buffering elements, power limiters, nonlinear enhancers in a magnon transistor, and components of logic gates (please check Ref. [75] for the corresponding references).

## *3.2. Two-dimensional structures*

The guiding of information carried by spin waves in two dimensions is required for the realization of magnonic circuits and is one of the important challenges modern magnonics is facing. As was shown in previous section, spin-wave dispersions are highly anisotropic in in-plane magnetised films and, in the simplest case of dipolar waves, BVMSWs and MSSWs exist in different frequency ranges (see case of a plane film in Figure 1) [2, 3]. Therefore, the realization of simple magnonic circuits of the type of printed circuit boards in electronics is not possible and alternative solutions should be used.

One likely possible way to achieve this was proposed in Ref. [175] with a usage of a T-shaped structure shown in Figure 5a. Counterpropagating MSSWs are excited by two microstrip antennas in a Permalloy ($Ni_{81}Fe_{19}$) strip (vertical waveguide in the Figure), which was saturated along its short axis by an externally applied magnetic field. In the center of this strip a second perpendicular strip was patterned to support the propagation of BVMSWs along the magnetization. Brillouin Light Scattering spectroscopy was employed to measure the spin-wave intensity in the structure [175]. In an unstructured film, the MSSW-to-BVMSW conversion is forbidden as was discussed above. However, in the waveguides the quantization of spin waves across the width of the waveguide leads to the modification and to the overlap of spin-wave dispersions (see Figure 1). Moreover, the internal field in the perpendicularly magnetized strip is smaller than the internal field in the longitudinally magnetized waveguide due to significant demagnetization. These factors allow for the coexistence of spin waves propagating parallel as well as perpendicular to the magnetization at the same frequency – see left panel in Figure 5a. As a result, the generation of BVMSW by the originally-excited MSSWs was shown experimentally and by means of numerical simulations [175]. In the right panels of Figure 5a, the simulated phase fronts of the propagating spin waves are shown with a colour code. In the top and bottom panels the spin waves are excited with the same and opposite phases correspondingly. The MSSWs of the same phases excite the $n=1$ symmetric BVMSW mode (top panel) while MSSWs of opposite phases excite the asymmetric $n=2$ BVMSW mode (bottom panel). These results were confirmed by phase-sensitive BLS measurements. The same type of mode conversion was also investigated for larger size YIG T-shaped spin-wave splitter in [178]. It was revealed that the spin wave beams in the outputs of the splitter are generally given by a superposition of both even and odd modes, with the details dictated by the dispersion overlap. By adjusting the frequency of the incident wave, it is possible to alter the character of the output beams or to switch the output off completely.

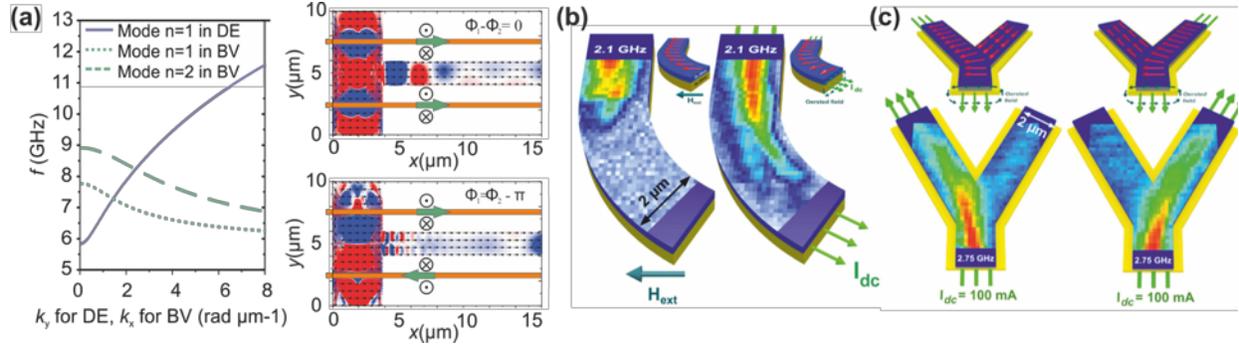

**Figure 5.** (a) Left panel: Spin wave dispersion relations, calculated for T-structure for the $n=1$ MSSW width mode, and for $n=1$ and $n=2$ BVMSW width modes [175]. Right panels: Snapshots of a simulation showing the basic principle of the investigated mode conversion (the color coding represents the out-of-plane component of the magnetization). The orange bars represent the positions of the simulated microwave antennas (in the top panel antennas are oscillating in phase, in the bottom panel they are in anti-phase). (b) Two-dimensional intensity distributions of spin waves excited with an externally applied magnetic field (left panel) and applied dc pulses with an amplitude of 66.7 mA (right panel) obtained with BLS spectroscopy [176]. (c) Two-dimensional BLS mapping of the spin-wave propagation path illustrating the switching: spin waves are guided through the Y-junction and only propagate in the same direction as the current flow. Red arrows in the insets shown the magnetization orientation [177].

A similar approach of the dispersion mismatch between the narrow magnonic waveguide and a wide antenna was used even earlier in Ref. [100] to excite spin waves by a global microwave field. This wireless method of spin-wave excitation uses uniform FMR-based antenna that couples to the microwave field and converts it into finite wavelength spin waves propagating in strip magnonic waveguides. The functionality of this method is demonstrated on the micrometer scale devices using time resolved scanning Kerr microscopy. This approach is especially important for the field of magnonics since these antennas can be placed at multiple positions on a magnonic chip and can be used to excite mutually coherent multiple spin waves for magnonic logic operations [99, 100].

Another way to guide spin waves in two dimensions was proposed by Vogt et al. in Ref. [176]. Spin-wave propagation in a Permalloy waveguide comprising an S-shaped bend was studied using BLS spectroscopy. In opposition to the cases discussed above, this method does not involve any conversion between the different modes. Instead, a non-uniform biasing field was used to magnetize the bent spin-wave waveguide transversally and to ensure the spin-wave propagation conditions for MSSW. For this, a direct current flowing through a gold wire underneath the Permalloy waveguide provided a local magnetic field – see Figure 5b. The mapped BLS intensity distribution of spin waves in the bent region of the waveguide is shown with color code in the Figure (the magnetization of the structure is shown in the insets with red arrows). The left panel shows that spin waves are not able to propagate through the bent structure if it is uniformly magnetized by an external field. As opposite, if a direct current is flowing through the bilayer (right panel in Figure 5b), the spin wave is guided within the curved waveguide. The advantage of this method is a possibility to control spin-wave propagation (in the simplest case to switch it on or off) and its drawback is the usage of electric current that generates parasitic Joule heat.

The same approach was used in Ref. [177] to guide spin waves in a Y-shaped structure – see Figure 5c. Electric current-induced locally generated magnetic fields, rather than uniform external fields, oriented the magnetization into only one arm of the Y-structure perpendicularly to the spin-wave propagation direction ensuring the propagation conditions for MSSWs. The investigated structure is proposed to be used as a spin-wave multiplexer (switch) that can be used e.g. for the guiding of spin-wave information to one or another magnonic data processing component. An interesting experimental finding in [177] is that the Y-structure is efficient for an angle of 60 degrees between the output arms and is much less efficient for angles of 30 and 90 degrees. One possible explanation is that the formation of caustic beams discussed in the next section takes place in the experiments.

### *3.3. Spin-wave caustics*

Strong spin-wave anisotropy in in-plane magnetized films discussed above is not necessarily a drawback for the transfer of information in two dimensions. For example, it opens an access to the guiding of spin waves even in an un-patterned film in the form of *spin-wave caustics* – non-diffractive wave beams with stable sub-wavelength transverse aperture [7-10]. The direction of these beams is controlled by the magnetic field and therefore caustics can selectively transfer

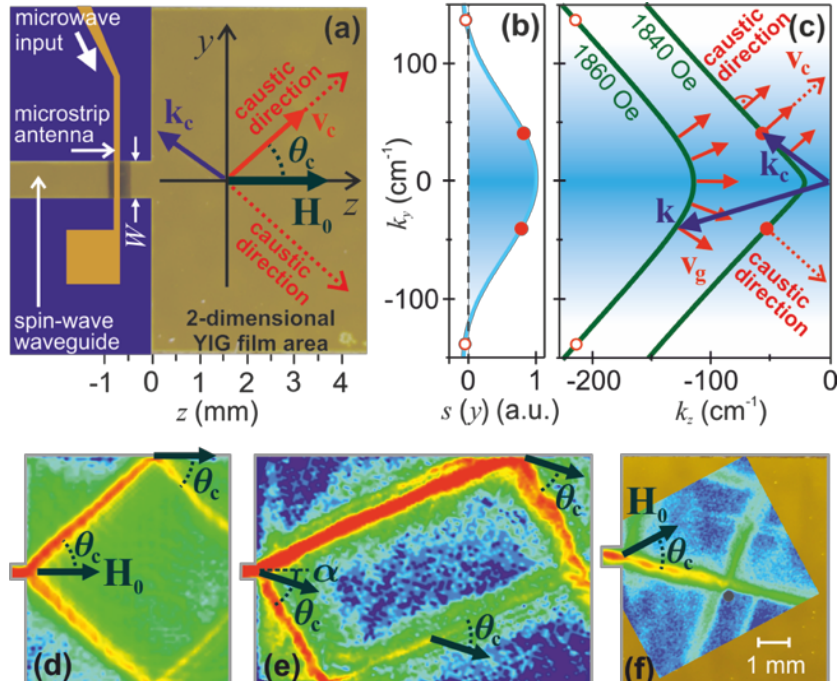

**Figure 6.** (a) YIG structure used for investigations of caustic spin-wave beams [7]. (b) Excitation amplitude of the waveguide antenna as a function of the transverse wave vector $k_y$. The shading marks the area were the spin-wave excitation is efficient. (c) Isofrequency curves for different magnetic fields. The filled circuits show caustic points. (d)-(f) Spin-wave intensity of the propagating caustic beams and their scattering from medium boundaries (d), (e) and a defect (black dot) (f) measured using BLS spectroscopy.

information to a spin-wave data processing element [19].

In an anisotropic medium, the direction of the wave group velocity $\vec{v}_g$ indicating the direction of energy propagation generally does not coincide with the direction of the wave vector $\vec{k}$. When the medium's anisotropy is sufficiently strong, the direction of the group velocity of the wave beam may become independent of the wave vectors of the waves forming the beam in the vicinity of a certain carrier wave vector $\vec{k}_c$. In such a case, wave packets excited with a broad angular spectrum of wave vectors (i.e. with a broad spectrum of phase velocities) may be channeled along this direction. The ideal source for such a wave packet is a point source whose size is comparable or smaller than the carrier wavelength of the excited wave packet. A simple excitation source delivering wave packets with wide angular spectra is shown in Figure 6a. A microstrip antenna was used to excite BVMSWs into a narrow spin-wave waveguide and the waves were then guided into a continuous area of a film. The transition between the waveguide and the continuous area of the film acts as a point source [7]. Figure 6b presents the $k_y$ spectrum of spin waves excited in such a way. The shaded area indicates where wave excitation is effective. The isofrequency curves of BVMSWs for two different magnetic fields are shown in Figure 6c. On the linear segment of the iso-frequency curve corresponding to 1840 Oe, the direction of $\vec{v}_g$ is the same for all wave vectors. Thus, a caustic wave beam can be formed, and the energy of this beam will propagate along the caustic directions perpendicular to the linear segments of the isofrequency curve making the angle $\theta_c$ with the magnetic field $\vec{H}_0$ – see the two beams measured experimentally using BLS spectroscopy in Figure 6d. The condition $d^2k_z/dk_y^2=0$ defines the carrier wave vector $\vec{k}_c$, the group velocity $\vec{v}_{gr}$, and the propagation direction $\theta_c$ of the caustic beam [7]. In contrast, for the larger magnetic field 1860 Oe the caustic points (open red circles in Figures 6b and 6c) are situated outside of the region of the effective excitation of the spin waves by the waveguide opening and therefore no caustics will be formed for this field [7].

Figure 6d demonstrates the scattering of caustic spin-wave beams from the YIG film boundaries when the field $\vec{H}_0$ is directed along these boundaries [7]. The boundary region, from which the propagating caustic wave beam scatters, acts as a secondary wave source, whose finite size is of the order of the beam's width. This secondary source also radiates a wave packet with a wide angular spectrum that again forms caustic wave beams propagating at the same angle $\theta_c$ to the anisotropy axis defined by $\vec{H}_0$ as the initial wave beam. The rotation of the bias magnetic field clockwise through the angle 20 degrees with respect to the medium boundary changes the pattern of the beam's reflection. The direction of the secondary, reradiated wave beam is determined by the direction of the inclined anisotropy axis in the medium $\vec{H}_0$ rather by the rule of reflection in linear optics – see Figure 6e. A similar effect is seen in Figure 6f, where the bias magnetic field was rotated counterclockwise by 30 degrees, to observe caustic beam scattering at the intentionally made defect in the film (shown as a black dot).

Besides YIG structures [7, 9], Permalloy [8] and Heusler materials [10] have been used to investigate spin-wave caustics. In the latter case, the nonlinear generation of higher harmonics leading to the emission of caustic spin-wave beams from localized edge modes was reported. The radiation frequencies of the propagating caustic waves were at twice and three times the excitation frequency [10]. In Ref. [179] a collapse of non-linear spin wave solitons and bullets was investigated experimentally and theoretically. It was shown that the collapse results in the generation of spin-wave caustic beams with the angles modified with respect to stationary caustics source due to the Doppler shift. It was demonstrated in Ref. [180] that the nonuniformity of the internal magnetic field and magnetization inherent to magnetic structures creates a medium of graded refractive index for propagating magnetostatic waves and can be used to steer their propagation. The two-dimensional diffraction pattern arising in the far-field region of a ferrite slab in the case of a plane wave with noncollinear group and phase velocities incident on a slit is investigated theoretically in Ref [181]. In Ref. [182] the authors have combined the dipole-dipole and Dzyaloshinskii-Moriya interactions that resulted in the formation of unidirectional caustic beams in the Damon-Eshbach geometry. Finally, a switchable spin-wave signal splitter based on the controllability of the caustic spin-wave beam direction by locally applied magnetic fields was recently demonstrated in Ref. [183].

## 3.4. Directional couplers

Several different concepts of magnonic logic and signal processing devices will be discussed later, but one of the unsolved problems of the magnonic technology is an effective and controllable crossing of magnonic conduits without interactions that is required for the realization of a functioning magnonic circuit. Unfortunately, a simple X-type crossing structure [146, 184] has a significant drawback, since the crossing point acts as a spin-wave re-emitter into all four connected spin-wave channels. The usage of a third dimension like it is done in electronics is also problematic due to the strong anisotropy of spin waves and demagnetizing effects. Thus, an alternative solution for the realization of spin-wave interconnections is necessary. One of the promising solution is based on the dipolar interaction between magnetic spin-wave waveguides. Originally, such a spin-wave coupling had been studied theoretically in a "sandwich-like" vertical structure consisting of two infinite films separated by a gap [185, 186]. However, the experimental studies of such structure are rather complicated due to the lack of access to the separate layers that is required for the excitation and detection of propagating spin waves. The configuration of a connector based on two laterally adjacent waveguides, which is well-studied in integrated optics, was recently proposed by Sadovnikov et al. also for magnonics [187] and has been studied experimentally using

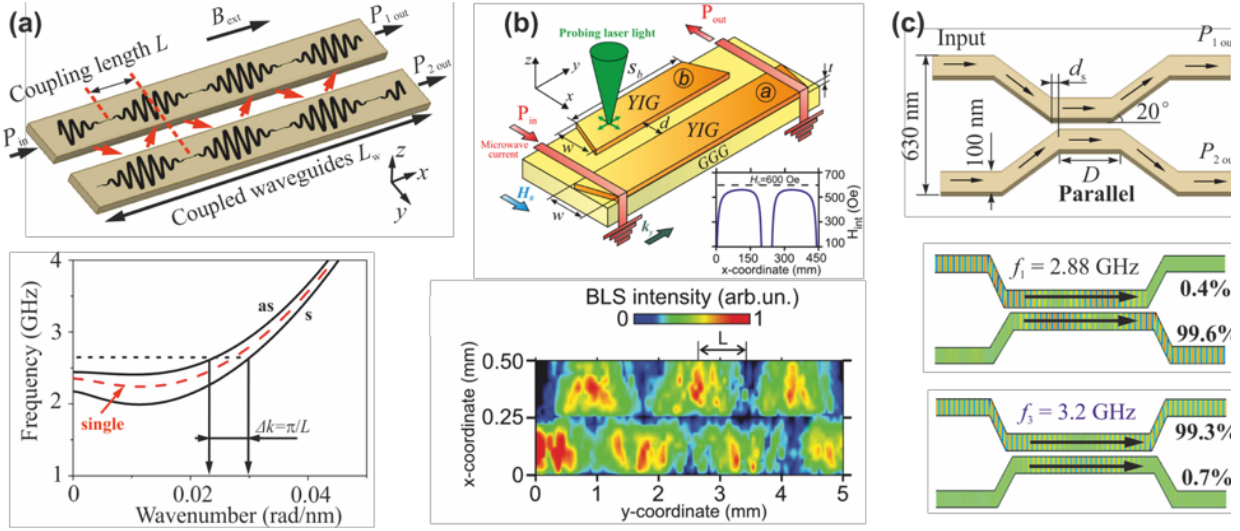

**Figure 7.** (a) The operational principle of a directional coupler based on two dipolarly coupled spin-wave waveguides [189]: Solid black lines illustrate the periodic energy exchange between the two interacting waveguides with a spatial periodicity of $2L$. Bottom panel: The red dashed line shows the dispersion characteristic of the lowest spin-wave width mode in an isolated single spin-wave waveguide. Solid black lines show the dispersion curves of the "symmetric" (s) and "anti-symmetric" (as) lowest collective spin-wave modes of a pair of dipolarly coupled waveguides [189]; (b) Top panel: Schematic of the structure under investigation in Ref. [187]. Inset shows profile of the static internal magnetic field along the $x$-axis. Bottom panel: Normalized color-coded BLS spin-wave intensity map. The lower waveguide is excited with a microwave antenna. (c) Top panel: Schematic view of the nano-scale directional coupler studied in [189]. The widths of the YIG waveguides are 100 nm, the thickness is 50 nm, the gap 30 nm. Bottom panels: Directional coupler as a crossing of magnonic conduits (middle pannel) or as a switchable transmission line (bottom panel). The numbers show the percentage of the spin-wave output energy in each arm.

Brillouin Light Scattering (BLS) spectroscopy for macroscopic YIG structures [157, 187, 188]. The downscaling of the same approach to the nanoscale was explored by means of numerical simulations in Ref. [189].

The general idea of the directional coupler is as follows [189]: In the case, where two identical magnetic strip-line spin-wave waveguides are placed sufficiently close to one another (see top panel in Figure 7a), the dipolar coupling between the waveguides leads to a splitting of the lowest width spin-wave mode of a single waveguide into the symmetric and anti-symmetric collective modes of the coupled waveguides - see bottom panel in Figure 7b. Thus, in a system of two dipolarly-coupled waveguides, two spin-wave modes at different wavenumbers $k_s$ and $k_{as}$ ($\Delta k = |k_s - k_{as}|$) will be excited simultaneously in both waveguides. The interference between these two propagating waveguide modes will lead to a periodic transfer of energy from one waveguide to the other as shown in the Figure 7a, so that the energy of a spin-wave excited in one of the waveguides will be transferred to the other waveguide after propagation over a certain distance that is defined as the coupling length $L=\pi/\Delta k_x$ [185]. The case which is of interest for applications is the one in which a spin wave is originally excited in only one waveguide (see Figures 7b and 7c). The output powers for both waveguides can be expressed as [185]: $P_{1out}=\cos^2\left[\pi L_w/(2L)\right]$ for the first waveguide, and $P_{2out}=\sin^2\left[\pi L_w/(2L)\right]$ for the second one, where $L_w$ is the length of the coupled waveguides, and $P_{in}$ is the input spin-wave power in the first waveguide. The dependence of the normalized output power of the first waveguide can be expressed as $P_{1out}/(P_{1out}+P_{2out})=\cos^2\left[\pi L_w/(2L)\right]$. Thus, the interplay between the length of the coupled waveguides $L_w$ and the coupling length $L$, which is strongly dependent on several external and internal parameters of the system, allows one to define the ratio between the spin-wave powers at the outputs of two coupled waveguides, and, thus, define the functionality of the investigated directional coupler.

The experimental demonstration of the spin-wave coupling in two laterally adjacent magnetic strips was performed in Ref. [187]. The sketch of the structure under investigations is shown in the upper panel of Figure 7b. Identical strips with a width of 200 μm and a thickness of 10 μm were separated by a gap of 40 μm. The magnetic strips have a trapezoidal form in order to minimize spin-wave reflection at their ends. The spin-wave intensity mapped with BLS spectroscopy (see bottom panel in Figure 7b) clearly shows the periodic transfer of spin-wave energy from one waveguide to another and back. It was shown that the coupling efficiency depends both on the geometry of the spin-wave waveguides and on the characteristics of the spin-wave modes [187]. In the following work, the Authors have improved the spin-wave coupling by optimizing the design and have investigated it for different types of MSWs [188]. Also, a nonlinear coupling regime of spin waves in adjacent magnonic crystals has been investigated experimentally in macroscopic waveguides [157]. It was shown experimentally as well as by using numerical simulation that a nonlinear phase shift of spin waves in the adjacent magnonic crystals leads to a nonlinear switching regime at the frequencies near the forbidden magnonic gap.

The dipolar coupling of nano- rather than macro-scale YIG spin-wave waveguides with parallel and anti-parallel orientations of the static magnetization along the long axis of the waveguides was studied using micromagnetic simulations in Ref. [189]. A general analytic theory describing the spin-wave coupling in the adjacent spin-wave waveguides is also developed in the same work. The coupling length $L$, over which the energy of the spin waves is transferred from one waveguide to the other, is studied as a function of the spin-wave wavenumber, the geometry of the coupler, the applied magnetic field, and the relative orientation of the static magnetization in the waveguides. Further, the design of the nano-scale directional coupler, in which strong shape anisotropy orients the static magnetization along the direction of the spin-wave propagation (see black arrows in top panel of Figure 7c) and ensures practically reflectionless spin-wave propagation (only few percent of spin-wave energy was reflected), was proposed [189]. The bottom panels in Figure 7c show (simulated spin-wave amplitudes are color coded) that the change in the spin-wave frequency leads to a change of the coupling length and, as a result, the spin wave can be guided to the second arm of the directional coupler (middle panel) or can be sent back to the same arm (bottom panel). In general, it is demonstrated that the directional coupler can be used as an element to cross waveguides (discussed in the beginning of the section, middle panel in Figure 7c), as a controlled multiplexer, a frequency separator, or as a power divider for microwave signals [189]. Moreover, the proposed device has the additional benefit of dynamic reconfigurability of its functionality within a few tens of nanoseconds. The nanometer sizes of the proposed directional coupler make it interesting and useful for the processing of both digital and analog microwave information.

## 4. Spin-wave excitation, amplification, and detection

Besides the transfer of information carried by spin waves, which is the main topic of the previous section, other important challenges the field of magnon spintronics is facing are the excitation, manipulation, amplification, and, finally, the detection of spin waves. The spin-wave manipulation was already partially discussed (e.g. in the part on magnonic crystals) and will be discussed in the section on magnon-based data processing. Often spin waves should be amplified in order to compensate spin-wave damping or to restore the intensity of the spin-wave signal after splitting into two channels. The state of the art microwave and spintronics approaches for the amplification of spin waves are discussed here.

The techniques used for the spin-wave excitation and detection in a laboratory have already been already overviewed in the section on materials and methodology in magnonics. However, the majority of these approaches (e.g. optical techniques) requires large complex equipment and cannot be implemented into magnonic devices. In this section, the selected techniques for the excitation and detection of spin waves that have the potential to be implemented into a magnonic chip are discussed. As opposite to all-magnon approaches, in which the electron-magnon conversion is not required (which will be discussed later) [24], the majority of computing-oriented spin-wave studies assumes that data should be coded from electric signal to magnons in each magnonic unit, should be processed and converted back to the electric signal afterwards. This will simplify the clocking of spin-wave devices [23] and will allow for compatibility with existing CMOS technology [20-23]. Therefore, the size of the convertors and their efficiency will define the sizes of future devices as well as their power consumption. High-potential microwave and spintronics approaches suitable for this task are discussed. Please note, that one more approach based on the excitation of spin waves by electric field using magneto-eclectic cells [21] will be discussed in the next section.

### 4.1. Parametric pumping

Different methods of spin-wave amplification were discussed e.g. in Ref. [4], but probably the one of the most important is parametric pumping approach [2]. A parametric process is one in which a temporally periodic variation in some system parameter affects oscillations or waves of another parameter in this system and can lead to their amplification. Perhaps the most well known of all parametric amplifiers is the child's swing. The phenomenon of parametric instability in spin-wave systems was discovered by Bloembergen and Damon in 1952, through the observation of a nonlinear spin-wave damping effect [190]. An explanation was later proposed by Anderson and Suhl [191] and a corresponding theory was developed. The theory is based on the consideration of interactions between spin-wave eigenmodes in a system, in particular between a uniform FMR mode $k=0$ and propagating spin-wave modes. When the correct conditions are fulfilled, energy from the driven uniform mode is pumped into the propagating modes, leading to their amplification from a thermal level. Since the alternating magnetization of the pumping mode in such a system is always perpendicular to the static bias field, this mechanism became known as spin-wave instability under *perpendicular pumping* [2]. The order of the parametric instability $m$ is defined simply by the law of conservation of energy $m\omega_p = \omega_1 + \omega_2$, where $\omega_p$ is the frequency of the pumping magnon, $\omega_1$ and $\omega_2$ are the frequency of the pumped (or secondary) magnons. Perpendicular parametric pumping is also often described in terms of multi-magnon scatterings. Thus, the instability of the first order $m=1$ is named three-magnon splitting and the instability of the second order $m=2$ is named four-magnon scattering.

A related but physically different phenomenon of *parallel pumping* was discovered a few years later by Schlömann, Green and Milano [192] and occurs when a spin wave receives energy from a double-frequency alternating magnetic field applied along the direction of magnetization. In this case, the magnetic field rather than an alternating magnetization is responsible for the effect and the back influence of the pumped spin waves on the pumping signal is usually ignored. In terms of energy quanta, parallel pumping can be understood as the creation of two magnons from a single pumping photon. The energy and momentum conservation laws for the instability of the spin wave under parallel pumping can be written as [2]:

$$\begin{aligned} \omega_p &= \omega_{m1} + \omega_{m2} \\ k_p &= k_{m1} + k_{m2} \end{aligned} \qquad (6)$$

where $\omega_{m1}$, $\omega_{m2}$ and $k_{m1}$, $k_{m2}$ are the frequencies and the wavenumbers of the interacting magnons, and $\omega_p$ and $k_p$, the frequency and wavenumber of the pumping photons. From the law of the conservation of energy, in the simplest case, the frequencies of the interacting magnons are both equal to half the pump frequency $\omega_{m1}=\omega_{m2}=\omega_p/2$ – see Figure 8a. The photon wavenumber is much smaller than that of the magnons $k_p \ll k_{m1}, k_{m2}$. Therefore, the interacting magnons are usually counter-propagating $k_{m1}=-k_{m2}$ [11, 12]. In order to enable interaction of the double-frequency $\omega_p$ pumping field with the magnetization precession of frequency $\omega_{m1}$, an alternative component of the magnetization oriented along the pumping field at frequency $2\omega_{m1}$ is required [103]. In the case of an in-plane magnetized magnetic thin film, this dynamic longitudinal component is non-zero as a consequence of the shape anisotropy, that results in an elliptic magnetization trajectory – see Figure 8b. Therefore, parallel pumping is efficient in thin films and the amplification of travelling MSSWs [193] and BVMSWs [194, 195] in YIG waveguides have been demonstrated. The gain of the amplification reached a value of 30 dB in some experiments. In all but a few special cases, parametric amplification of magnons in mesoscopic samples must be performed in a pulsed, rather than continuous, pumping regime. The reason for this is that when pumping is applied, the amplitudes of thermal exchange modes, which are degenerate with the magnetostatic magnons of interest, start to grow exponentially [196]. Taking into account that these thermal waves (often named dominant modes) usually exhibit a lower damping, their amplification rate is larger which allows their amplitudes to overcome the amplitude of the signal-carrying spin wave quite fast in spite of their small original amplitudes [196]. In order to avoid undesirable competition between these modes and the amplified spin-wave signal packet, the pumping duration is usually considerably shorter than the characteristic relaxation time of the spin wave. The most comprehensive theory that describes parametric instabilities as well as the interactions between the magnons is named *S-theory* and was developed by L'vov and Zakharov [46, 197].

The amplification of nonlinear eigen-excitations of magnetic media, namely, spin-wave envelope solitons [198] and bullets [12], constitutes a very interesting separate problem since the ratio between the length and the amplitude of a soliton is fixed. The solution was realized by the usage of localized pulsed parametric pumping and an amplification of 17 dB of fundamental BVMSW solitons has been demonstrated by Melkov et al. [198]. Moreover, according to the momentum conservation shown in Eq. (6), the secondary wave of a wave vector $-k_{m2}$ propagates in the opposite direction and is phase-conjugated. Therefore, the Wave Front Reversal (WFR) of linear [11] as well as non-linear [12] spin-wave packets were realized.

Finally, parallel parametric pumping is a very efficient mechanism to excite spin waves of arbitrary wavelengths and, is therefore usually used for the magnon injection (namely amplification of spin waves from thermal level) in the experiments on Bose-Einstein Condensation (BEC) of magnons [13-16]. Using parametric pumping, a pumped magnon density of $10^{18}$– $10^{19}$ cm$^{-3}$ can be reached [13]. Although this density is much smaller than that of thermal magnons at room temperature ($10^{21}$–$10^{22}$ cm$^{-3}$), it is sufficient to increase the chemical potential of the magnons to the energy of the lowest magnon state even at the room temperature. As a result, the formation of Bose–Einstein Condensate (BEC) of magnons was reported by Demokritov et al. in Ref. [13]. Later, Serga et al. [15] have demonstrated that parametric pumping can create remarkably high effective temperatures in a narrow spectral region of the lowest energy states in a magnon gas that results in strikingly unexpected transitional dynamics of a Bose–Einstein magnon condensate: The density of the condensate increases immediately after the external magnon flow is switched off (evaporative supercooling mechanism [15]). Finally, the first evidence of the formation of a room-temperature magnon supercurrent was reported recently by Bozhko et al. [16] (the group of B. Hillebrands). The appearance of the supercurrent, which is driven by a thermally-induced phase shift in the condensate wavefunction, is evidenced by analysis of the temporal evolution of the magnon density measured by means of BLS spectroscopy.

Majority of the experiments described above were performed using YIG samples of micrometer thicknesses and macroscopic lateral dimensions. An important breakthrough in the utilization of the parallel parametric pumping at the micrometer-scale was performed in a set of papers by Brächer et al. [102, 199-201] (see also review [103]). Parallel parametric generation of spin waves was studied by means of BLS spectroscopy in a longitudinally magnetized Permalloy magnonic waveguide of 2.2 µm width [200]. A 1.2 µm-wide microstrip antenna was placed perpendicularly on top of the spin-wave waveguide to apply the double-frequency pumping magnetic field oriented parallel to the magnetization of the waveguide. First, the parametric instability was investigated in the absence of any coherently-excited spin waves [200]. Taking into account the quite large wavenumber range of micro-focused BLS spectroscopy (approximately $k_{max}^{BLS} \approx 20$ rad/µm), this allows for a direct mapping of the parametrically excited dominant spin-wave mode exhibiting the smallest threshold of the parametric instability. The analysis of the spatial distributions of the generated spin waves have shown that odd as well as even BVMSW waveguide modes can be excited parametrically. Furthermore, it was revealed that the generation process takes place underneath the antenna where the pumping field is at its maximum only due to the threshold nature of the parametric instability (the magnetic field away from the antenna is too small to reach the threshold of the parametric instability). Parametric amplification of propagating spin waves was proven by the investigation of the spatial decays of spin waves in Ref. [102]. This time, the pumping field was applied by a microwave current flowing through a copper microstrip line underneath the waveguide. It was shown that amplified spin waves propagate distances of about 30 μm and an amplification factor of approximately 10 dB was achieved. To avoid mode competition with a dominant spin-wave mode and saturation of the amplification, short pulses with pumping powers close to the threshold power of parametric generation were used [102].

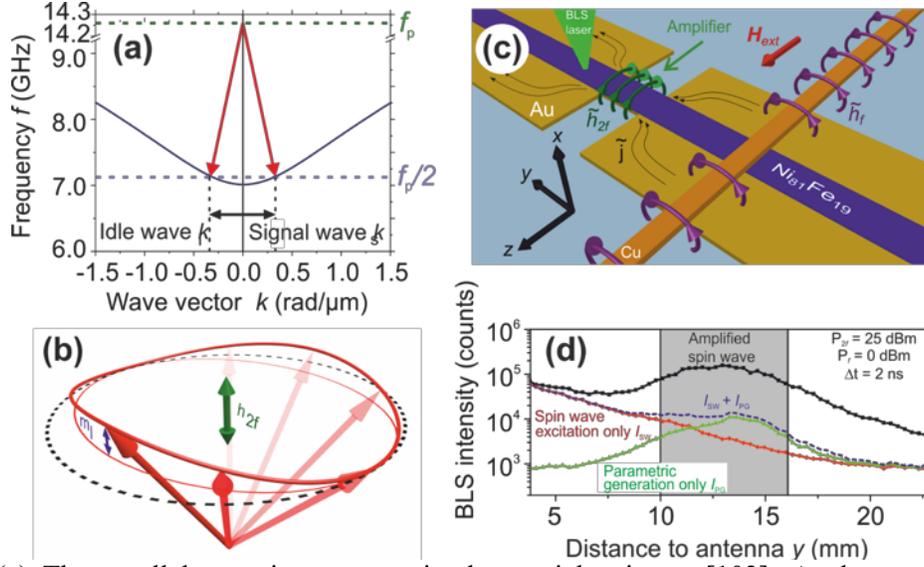

**Figure 8**. (a) The parallel pumping process in the particle picture [103]: A photon with a small wavevector splits into a pair of counter-propagating magnons, leading to the formation of the signal and idler wave at one-half of the photon frequency. The solid dark blue line represents the spin-wave dispersion of the fundamental waveguide mode. (b) Schematic of the elliptical magnetization trajectory in a thin film [103]. This trajectory gives rise to the double-frequency longitudinal dynamic magnetization component $m_l$ that interacts with the microwave pumping field $h_{2f}$. (c) Sample layout for the localized parallel parametric amplification studied by BLS microscopy [199]. (d) Space-resolved BLS intensity arising from the parametric amplification of the externally excited spin waves (black squares), from the antenna excitation only (red dots) and from parametric generation only (green triangles). The shading marks the position and the length of the amplification region.

The utilization of a localized rather than uniform parametric pumping in a transversally magnetized in-plane magnetized Permalloy waveguide was demonstrated in Refs. [199, 201]. The localization was realized by combining the threshold character of parametric generation with a spatially confined enhancement of the amplifying microwave field by introducing a narrowed region in a microstrip transmission line – see Figure 8c. The Figure 8d shows the functionality of the localized spin-wave amplifier. Spin waves are excited at the antenna by a 15 ns long microwave pulse with a carrier frequency of 6 GHz and decay exponentially if no pumping signal is applied (see the red line in the Figure). The generation of the dominant spin wave only due to the application of the parametric pumping signal at 12 GHz is shown with the green line. Finally, the amplified spin wave is shown by the black line in Figure 8d. An amplification of the propagating wave is clearly visible. Moreover, it was shown by time-resolved measurements that the amplification is efficient as long as the pumping is timed properly with respect to the arrival time of the spin-wave packet. In the case of a strong pumping, this timing is crucial and the spin waves have to arrive prior to the pumping pulse. If the applied pumping is rather weak, the timing becomes less important, a higher gain can be achieved, and the amplifications is practically independent of the arrival time of the spin-wave packet [199].

## *4.2. Spin Hall effect and spin transfer torque*

As was discussed above, the combination of magnonic devices with electronic circuits requires efficient means for magnon excitation by a charge current. Although magnons can be injected relatively easily by an AC current (e.g. using antenna structures), it is a quite complex problem if a DC current is used. One of the promising solutions is the usage of the *Spin-Transfer Torque* (STT) effect. In 1996 Slonczewski [111] and Berger [112] have predicted independently that the injection of a spin-polarized current in a magnetic metallic film can generate a Spin Transfer Torque strong enough to reorient the magnetization or to excite magnetization precession [202] in this film. In order to generate the spin-polarized current, the charge current is sent through an additional magnetic layer with a fixed magnetization direction. A device especially designed to excite the magnetization precession is named Spin-Torque Nano-Oscillator (STNO). A first microwave measurement of a spin-torque-driven precession was presented in 1998 by Tsoi et al. [203]. Krivorotov et al. demonstrated experimentally that STT can be used to control the magnetic damping and for the magnetization reversal of a nanomagnet [113].

The excitation of spin waves by STNO was observed by Demidov et al. in 2010 [114]. The authors used BLS spectroscopy in order to perform a two-dimensional mapping of waves emitted by the STNO into an in-plane magnetized Permalloy film. It was reported that the emission is directional and depends on the orientation of the applied magnetic field. However, since the propagation length of the emitted waves did not exceed one micrometer, the propagating character of the waves was not proven. Another report on spin-wave generation by STNO was presented by Madami et al. [115]. The authors used a normally magnetized Permalloy film, which was probed by BLS spectroscopy – see left panel in Figure 9a. In this case, the radial emission of spin waves propagating over a distance of a few micrometers was observed (see right panel in the Figure 9b). The experimentally obtained magnon free path agrees well with a theoretical estimation.

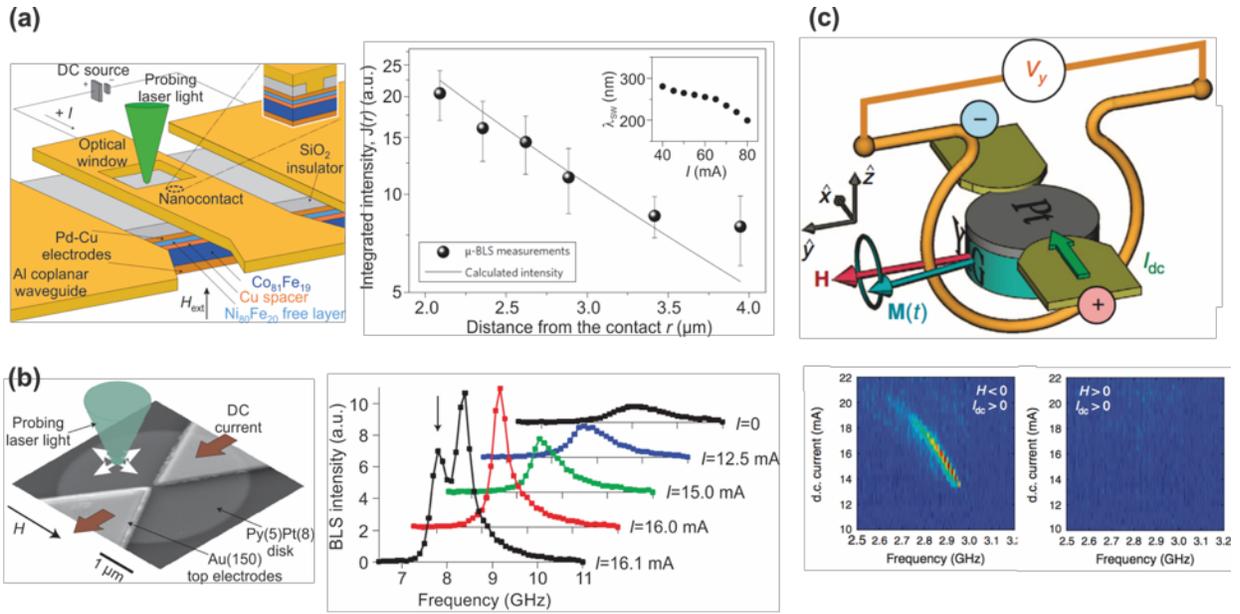

**Figure 9**. Magnon excitation by spin transfer torque. (a) Sample layout for studies of propagating spin waves induced by the spin-transfer torque (STT) [115]. An aluminium coplanar waveguide is deposited onto the spin-valve structure, and an optical window for BLS probing is etched into the central conductor of the waveguide close to the STT nano-contact. Right panel: Measured (symbols) and calculated (line) spin-wave intensity as a function of the distance from the center of the point contact. The inset shows the dependence of the spin-wave wavelength on the applied electric current. (b) Experimental layout of a magnetic nano-oscillator driven by a spin current generated by the SHE [116]. The device comprises a Pt (8 nm)/Py (5 nm) disk of 4 μm diameter with two Au electrodes separated by a 100 nm gap. Right panel: BLS spectra of the thermal spin-wave fluctuations at electric currents below the onset of the auto-oscillation. The spin current induced auto-oscillation peak (marked with a vertical arrow) appears at a current of 16.1 mA. (c) Sketch of the measurement configuration a device with two YIG/Pt microdisc (cale bar is 50 μm) [117]. The bias field is oriented transversely to the electric current flowing in Pt. The inductive voltage produced in the antenna by the SHE-STT generated precession of the YIG magnetization is amplified and monitored by a spectrum analyser. Bottom panels: power spectral density maps measured on a 4 μm YIG/Pt disc at fixed magnetic field and variable direction of electric current. Magnetization precession amplitude is color coded.

Another way in which spin-polarized electron current can be generated is based on the *Spin*

*Hall Effect* (SHE) caused by spindependent scattering of electrons in a non-magnetic metal or semiconductor with large spin-orbit interaction [204, 205]. Electrons flowing in a film, magnetized in-plane and transversely to the direction of the current, scatter with spin-asymmetry generating a spin current perpendicular to the film plane. This current, crossing the interface to an attached magnetic layer, generates a STT in this layer. A typical metal used in SHE studies is Pt due to its large conversion factor of charge current to spin current [205, 206]. Its thickness varies between 2 and 10 nanometers; these values are close to the spin-diffusion length in Pt [208, 125].

A great advantage of the SHE as a spin-current source is that a STT can be injected not only into a single local object (the diameter of a typical STNO is less than 100 nm) but, into a large area of a magnetic film. This allows for the realization of spin-wave amplification due to damping compensation. The first experimental observation of SHE-induced damping reduction was reported by Ando et al. in [206] in a Py/Pt bilayer. In the studies in Ref. [207], the variation of the damping parameter by a factor of four was demonstrated in Py/Cu/Pt multilayers studied by means of BLS spectroscopy. However, no spin-wave generation was achieved in these experiments, the probable reason being the strong nonlinear redistribution of the injected energy between many magnon modes. In subsequent studies, a modified design of the current-conducting structure containing bowtie-shaped electrodes with a 100 nm gap between them was used [116]. According to the model proposed by the authors, these electrodes allow for an increase in the density of the electric current applied to the Pt layer and for introducing a controlled radiation loss mechanism for the parasitic magnon modes previously disturbing the generation process. Consequently, the injected energy was concentrated onto the small area between the electrodes, and a single bullet-like spatially localized magnon mode was observed. Subsequently, the coherency of these magnons was proven by Liu et al. through microwave measurements [209]. Later, Duan et al. [210] demonstrated microwave oscillations of the magnetization in a ferromagnetic nanowire, where the geometric confinement dilutes the magnon spectrum and, thus, suppresses the parasitic nonlinear energy redistribution.

Another important advantage of the SHE-based STT is that no electric current needs to flow across the magnetic layer and, thus, the usage of a low damping magnetic dielectric material such as YIG is possible. In the pioneering work of Kajiwara et al. [124] the transmission of a continuous electric signal through a YIG film was demonstrated. For that, the SHE-based STT was used in order to convert an electric current into travelling spin waves. However, the exact conditions required for such conversion [79, 211, 212] or even for the SHE-STT based damping compensation [213, 214] were not defined. In this context, special attention has been attracted by the work of Hamadeh et al. where bowtie-like electrodes (see Figure 9b) were used for the damping compensation in YIG discs [215]. The decrease of the ferromagnetic resonance linewidth (which is a measure of the magnetic damping) by a factor of three was reported. Another approach was reported by Lauer et al. in Ref. [118] where the threshold of the parametric instability measured by BLS spectroscopy was used to determine the degree of STT-SHE controlled spin-wave damping. A macroscopically sized YIG(100 nm)/Pt(10 nm) bilayer of $4\times 2$ mm$^2$ lateral dimensions and pulsed current regime were under investigations rather than microscale structures with a continuous current. A variation in the relaxation frequency of ±7.5% was achieved for an applied current density of 5 10$^{10}$ A/m$^2$ depending on current polarity [118]. The SHE-STT amplification of propagating spin waves was studied experimentally in microscopic waveguides based on the nanometer-thick YIG/ Pt bilayers in Ref. [216]. It was shown that the propagation length of the spin waves in such systems can be increased by nearly a factor of 10. It was also demonstrated that, in the regime where the magnetic damping is completely compensated by the SHE-STT, the amplification of the spin wave was suppressed by the nonlinear scattering of the coherent spin waves from current-induced excitations [216]. An important breakthrough in the field was the demonstration of the SHE-STT excitation and transport of diffusive spin waves by Cornelissen et al. in Ref. [217]. It was shown experimentally that magnons can be excited and fully detected electrically and can transport spin angular momentum in YIG over distances of 40 μm. Gönnenwein et al. reported shortly afterwards in Ref. [218] on the observation of the same phenomenon due to investigations of a local and non-local magnetoresistance of thin Pt strips deposited onto YIG. Later, this phenomenon was used to build a spin-wave majority gate using a multi-terminal YIG/Pt nanostructure [219].

The SHE-STT generation of coherent spin-wave modes in YIG microdiscs by spin–orbit torque was reported by Collet et al. in Ref. [117]. Magnetic micro-discs with diameters of 2 and 4 μm were fabricated based on a hybrid YIG(20 nm)/Pt(8 nm) bilayer and were measured using a microwave antenna around the discs – see top panel in Figure 9c. Color plots of the inductive signal measured as a function of the relative polarities of magnetic field are presented in the bottom panels of Figure 9c. An auto-oscillation signal is clearly visible if the current and field polarities are properly chosen in accordance with the symmetry of SHE [117]. Further investigations of the phenomena using spatially-resolved BLS spectroscopy were reported in Ref. [220]. It was show that SHE-STT excited spin-wave modes exhibit nonlinear self-broadening preventing the formation of the self-localized magnetic bullet, which plays a crucial role in the stabilization of the single-mode magnetization oscillations. Time-resolved BLS spectroscopy measurements of the SHE-STT excited magnetization precession in YIG/Pt micro-discs was reported in [119]. It was shown that the magnetization precession intensity saturates within a time frame of 20 ns or longer, depending on the current density.

It was also demonstrated that a spin current and consequently, the STT may be induced by a thermal gradient rather than by an electric current and the SHE effect [221-223]. Particularly, recently the generation of coherent spin waves in YIG/Pt nano-wires by the application of thermal gradients across the structure was demonstrated in Ref. [224]. The nanowires were investigated at low temperatures by means of electrical measurements. It was shown that the Spin Seebeck Effect (SSE) induced spin current induced by the Ohmic heating of Pt is the dominant drive of auto-oscillations of the YIG magnetization. The magnon generation was also observed in Ref. [225] in a similar structure using time- and space-resolved BLS spectroscopy at room temperature in a pulsed heating regime. It was found that in this experiment the role of SSE is vanishing with respect to the heating of the Pt/YIG structure by the DC current pulses. The heating and consequent fast cooling of the structure resulted in the increase of the chemical potential of magnons and lead to the formation of Bose-Einstein Condensation. All these approaches open the door for the effective conversion of waste heat in future magnonic devices into spin waves or electric currents for further data transfer and processing [131, 226].

## 4.3. *Spin pumping and inverse spin Hall effect*

The field magnon spintronics assumes that after information is processed within a magnonic system it needs to be converted back to electronic signals. A conventional way to do this is to use a strip or a coplanar antenna, in which spin waves induce an AC current that is in turn rectified by a semiconductor diode. Another, recently discovered way is based on the combination of two physical effects: the *Spin Pumping* (SP) and the *Inverse Spin Hall Effect* (ISHE). In 2002, Tserkovnyak, Brataas, Bauer [121] showed theoretically that magnetization precession in a magnetic film will generate a spin-polarized electric current in an attached non-magnetic metallic layer. This process manifests itself in the increase in damping of a magnonic system [121, 227, 228]. The electrical detection of the spin pumping induced spin current was reported by Costache et al. in 2006 [122]. In the same year, Saitoh et al. [123] reported the observation of spin pumping using the ISHE. This effect refers to the generation of a charge current in a nonmagnetic metal by a spin current and is the reciprocal effect of the SHE. Due to this effect, a spin current induced in a Pt film by a precessing magnetization in an adjacent magnetic film is converted into a detectable DC voltage (see, e.g., the review by Hoffman [125]). Since then, the combined SP-ISHE mechanism is used as a convenient detection mechanism of magnons.

The SP-ISHE mechanism allows for measurements of the magnetization precession not only in metallic but also in YIG-based structures. A first report from Kajiwara et al. on the observation of an ISHE voltage in YIG/Pt bilayers [124] was followed by comprehensive studies of this phenomenon. The dependencies on the thicknesses of the nonmagnetic metal [208, 211, 229] and the YIG [230, 231] layer, as well as on the applied microwave power [231] have been reported. The influence of the interface conditions on the spin pumping efficiency [232, 233] was revealed, and the contributions to the SP effect by different spin-wave modes were studied [234]. An important milestone was the successful implementation of the combined SP-ISHE mechanism for the detection of propagating spin waves [235]. A typical geometry is shown in Figure 10a. A spin-wave detector in form of a 200 μm wide Pt strip was placed 3 mm away from the microstrip antenna. The Pt strip was used for the detection of the time-resolve ISHE voltage, as well as, for reference, as a conventional inductive probe. In the bottom panels of Figure 10a, both the AC and the electric signals produced by a 50 ns-long spin-wave pulse are shown. One can clearly see that the ISHE voltage appears with a delay of 200 ns determined by the spin-wave propagation time between the antenna and the detector. The ISHE nature of the electric signal is proven by the fact that the inversion of the direction of the biasing magnetic field results in the switching of the voltage polarity [236] – see the Figure. In the same experiment, it was also demonstrated that the spin-pumping efficiency does not depend on the spin-wave wavelength. d'Allivy Kelly et al. have used a similar experimental setup to demonstrate the ISHE detection of propagating magnons in a nanometer-thick YIG film [79]. Magnetostatic surface spin waves show nonreciprocal behaviour. The propagation direction of these waves can be reversed by a change in the polarity of the bias magnetic field [96]. An electric probe was used to measure the ISHE voltage at different points of a sample in Ref. [237].

It was shown that the ISHE voltage induced by the MSSWs depends on the field orientation. The combined SP-ISHE mechanism opened doors for the access to short-wavelength exchange magnons [43, 45]. The short-wavelength regime is of particular interest for nano-size magnonic applications. In order to excite exchange magnons, the parallel parametric pumping technique discussed above was used [43] - see Figure 10b. By the variation of the bias magnetic field, the magnon spectrum is shifted upward or downward to tune the wavelength of the magnon (see inner panel in the Figure). The spin pumping induced ISHE voltage is shown in the bottom panel of Figure 10b. One can see that magnons effectively contribute to the spin pumping within a wide range of wavelengths (down to 100 nm in Ref. [43]). Follow-up studies by Kurebayashi et al. [238], where parallel and perpendicular parametric pumping techniques were used for the magnon injection, have evidenced that the spin pumping efficiency is independent of the magnon wavelength within experimental error.

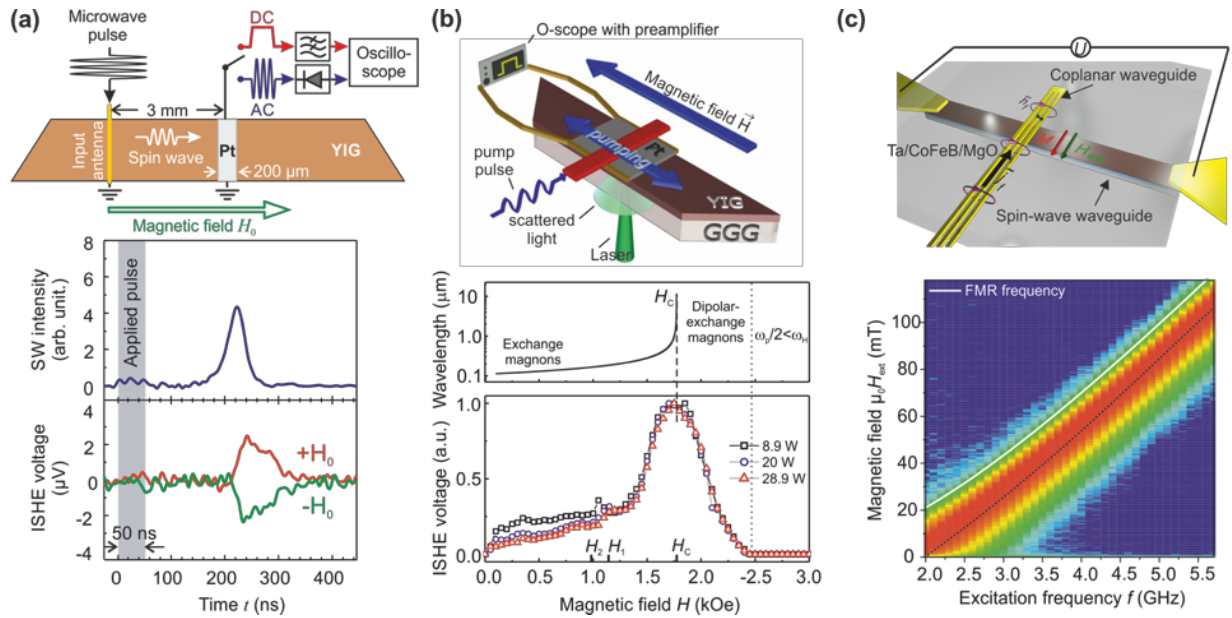

**Figure 10.** (a) Schematic illustration of the experimental setup for the Inverse Spin Hall Effect detection of propagating spin waves [235]. A spin-wave packet is excited in the YIG conduit using a microstrip antenna and is detected by the Pt strip placed 3 mm apart as AC and DC signals. Temporal evolutions of the spin-wave intensity (AC signal) and the ISHE voltage (DC signal) generated by a 50 ns-long spin-wave pulse are shown in the bottom panel for different field polarities. (b) Sketch of the experimental setup for the investigation of spin pumping by parametrically injected exchange magnons [43]. A 10 nm thick, 3×3 mm² Pt layer is deposited onto the 2.1 μm-thick YIG film. A microwave pumping field is applied using a microstrip antenna. Bottom panel: Normalized dependencies of the ISHE voltage induced by the injected magnons as a function of the bias magnetic field for different pumping powers. The calculated wavelength of the parametrically injected magnons is shown in the middle panel. (c) Schematic of the sample and the geometry of the spin-wave rectification measurements in [45]. A Ta/CoFeB/MgO trilayer is patterned into a spin-wave waveguide with leads to measure the ISHE voltage along the waveguide. A nanoscale coplanar waveguide antenna is placed on top to excite spin waves. Bottom panel: Color-coded measured ISHE voltage as a function of the applied frequency and applied magnetic field.

It was recently shown by Brächer et al. in Ref. [45] that exchange spin waves coherently excited by a microstrip antenna rather than by parametric pumping can be efficiently detected in a Ta/Co$_8$Fe$_{72}$B$_{20}$/MgO microscaled waveguide – see top panel in Figure 10.c. This layer system features large spin orbit torques and a large Perpendicular Magnetic Anisotropy (PMA) constant. The short-wavelength spin waves were excited by nano-scale coplanar waveguides (the smallest feature size was 50 nm) and were detected by way of SP-ISHE voltage as well as using micro-focused BLS spectroscopy [45]. The bottom panel in Figure 10.c shows the measured SP-ISHE spectra of the excited spin waves. The measured ISHE voltage is shown color-coded as a function of the applied magnetic field and frequency. It is demonstrated that the noise-limited maximum detectable wave-vector was about 42 rad μm$^{-1}$ (wavelength is around 150 nm) and is determined by the Fourier spectrum of the excitation source. These short wavelengths are not detectable by e.g. BLS spectroscopy, which is only sensitive to magnons down to about 300 nm wavelengths. Moreover, it was demonstrated that the spin-wave emission by the CPWs exhibits a strongly preferred emission direction due to the large PMA in the investigated spin-wave waveguide [45].

## *4.4. Magneto-electric cells and their usage for spin-wave logic*

For more than four decades the successful scaling of Complementary Metal–Oxide–Semiconductor (CMOS) Field-Effect Transistors (FETs) took place according to Moore's law [239]. The projection of scaling limits and quantum limits on the size of electronic transistors [240] stimulates a search for novel *beyond-CMOS* technologies. The benchmarking of these novel approaches became an important effort and it was nicely reviewed by Nikonov and Young in Refs. [20]. The authors have analyzed 11 new devices operating with three types of new magnetization switching mechanisms: (*i*) Spin Hall Effect, (*ii*) ferroelectric switching, and (*iii*) piezoelectric switching. The Spin Hall Effect based approaches were already described above in details. Ferroelectric devices rely on electric polarization and are very promising [241-243]. In this section, however, the piezoelectronic devices, which utilize stress and strain mechanisms to switch magnetization, are described. Since excitation, detection and manipulation of spin waves in these devices is performed using electric fields rather than currents, this approach has a large-potential for future spin-wave computing with low-energy consumption.

A feasibility study of logic circuits utilizing spin waves controlled by electric field for information transmission and processing by Khitun et al. was discussed in Ref. [244]. In the following publications [144], the authors proposed basic elements that include voltage-to-spin-wave and spin-wave-to-voltage converters, spin-wave waveguides, a spin-wave modulator, and a *Magneto-Electric (ME) cell*. The performance of the basic elements was demonstrated by experimental data as well as with the results of numerical modeling. The combination of the basic elements allowed for the construction of magnetic circuits for NOT and majority logic gates as well as for AND, OR, NAND, and NOR logic gates [21, 144]. The proposed ME cell requires special attention and is shown schematically in Figure 11a. It consists of piezoelectric and magnetostrictive films. The ME coupling occurs via strain, or in the case of high frequency, via acoustic waves. As the coupling between voltage and strain in piezoelectric films can be efficient [27], the ME cells are very promising as transducers with a high efficiency as it relates to energy along with a large output signal. An additional advantage of ME-cell-based transducers is their high scalability due to the absence of delocalized magnetic fields inherent to microstrip antennas and current-driven STT elements. A numerical modeling of the ME cells and a concept of magnetic logic circuits engineering is given in Ref. [245]. The utilization of ME cells to control the phase of spin waves was presented in Ref. [246].

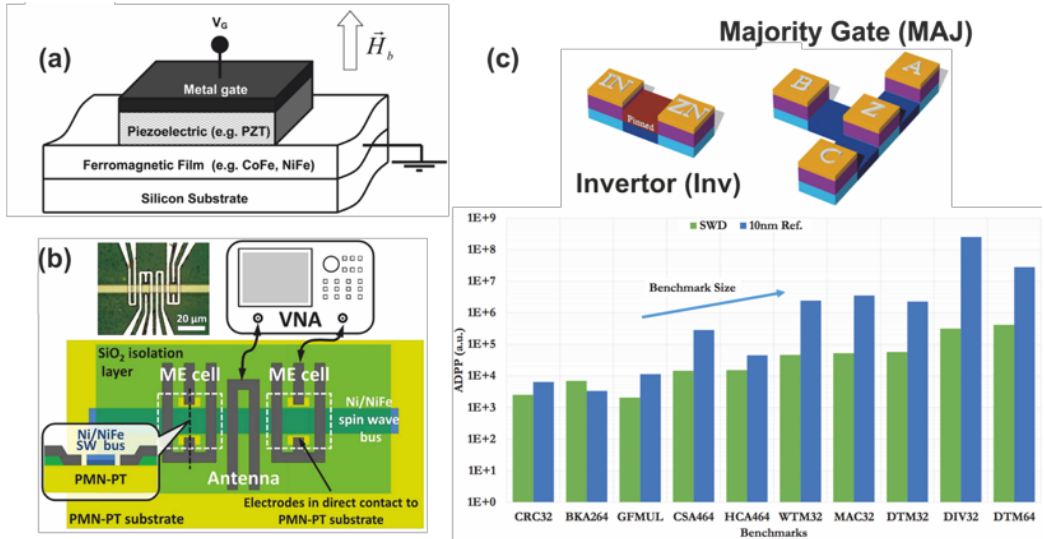

**Figure 11**. (a) Schematic view of the magneto-electric (ME) cell proposed in [144]. (b) Schematic of the experimentally realized ME Cell [247]. Spin wave generation and propagation is measured using a vector network analyzer. The inset shows a cross-section view of the ME cell. (c) Top panel: Gate primitives used for SWD circuits [22]. Bottom panel: ADP product of all benchmarks for spin-wave devices (green columns) and for the reference 10 nm CMOS (blue columns).

Cherepov et al. reported in Ref. [247] on the spin-wave generation by multiferroic magnetoelectric (ME) cell transducers driven by an alternating voltage. A multiferroic element consisting of a magnetostrictive Ni film and a piezoelectric $[Pb(Mg_{1/3}Nb_{2/3})O_3]_{(1-x)}$–$[PbTiO_3]_x$

substrate was used for this purpose – see Figure 11b. By applying an AC voltage to the piezoelectric, an oscillating electric field was created within the piezoelectric material and resulted in an alternating strain-induced magnetic anisotropy in the magnetostrictive Ni layer. The resulting anisotropy-driven magnetization oscillations propagate in the form of spin waves along a 5 µm wide Ni/NiFe waveguide. The authors have noted that the amplitude of the generated spin waves was rather low in this demonstration but, can be improved by using materials with lower damping or by geometrical and materials optimization [247].

Dutta et al. [23] proposed a comprehensive scheme for building a clocked non-volatile spin-wave device by introducing an ME cell that translates information from the electrical domain to the spin domain, magneto-electric spin-wave repeaters that operate in three different regimes (spin wave transmitter, non-volatile memory and spin wave detector), and a novel clocking scheme that ensures the sequential transmission of information and nonreciprocity. The authors have demonstrated that the proposed device satisfies the five essential requirements for logic application: nonlinearity, amplification, concatenability, feedback prevention and a complete set of Boolean operations [23]. Finally, Zografos et al. presented [22] a design and benchmarking methodology of Spin-Wave Device (SWD) circuits based on micromagnetic modeling. Spin-wave device technology is compared against a 10nm FinFET CMOS technology, considering the key metrics of area, delay and power. ME cell based spin-wave invertors and majority gates (see top panel in Figure 11.c) have been considered as primitives of complex SWD circuits processing up to a few hundreds of bits. Figure 11c shows the Area-Delay-Power Product (ADPP) which depicts how much the SWD circuits outperform the 10nm CMOS reference ones. On average, the area of future SWD circuits is expected to be 3.5 times smaller and the power consumption up to 100 times lower when compared with the 10nm CMOS reference circuits [22]. Therefore SWD appears as a strong contender for ultra-low power applications.

## *5. Conclusions and Outlook*

To conclude, magnonics and magnon spintronics are very active and promising fields with a number of break-through developments towards application in data processing that have already been demonstrated or are expected to be demonstrated in the near future. In order to clarify this statement, the concluding chapter is prepared in the form of brief discussions at a very basic level.

***Transport of spin-wave carried data.*** Data coded into spin-wave amplitude or phase is typically guided with the use of waveguides in the form of strips made of magnetic materials [19, 41, 54, 72]. Besides, spin-wave physics paves the ways to novel approaches for data transport e.g. in nonpatterned plane films in the form of caustics [7-10] or using dipolarly coupled waveguiding structures [187, 189]. A guiding of spin waves in two dimensions is presented in the section on magnonic circuits. The main characteristics of spin waves are given in the section on spin waves in thin-film waveguides.

***Types of signals processed using spin waves.*** Classically, spin waves are investigated with the view on operations with analogue information in the GHz frequency range (microwave filters, delay lines, amplifiers, etc.) [33-35, 152-155]. Also binary digital data can be coded into spin-wave amplitude or phase and conventional logic gates operating with the use of spin waves have been developed – see section on magnon-based processing of digital data. Alternately, novel computing concepts involving reversible computing [28, 139], neural networks [29], and quantum computing [251] also have high potential and the field of magnonics is suitable to realize them. Chapter 19 of Spintronics Handbook: Spin Transport and Magnetism, Second Edition, edited by E. Y. Tsymbal and I. Žutić (CRC Press, Boca Raton, Florida) (volume 3) discusses more magnonics concepts for Non-Boolean computing including holographic memory, pattern recognition, and prime factorization problem.

***Advantages for data processing proposed by spin waves.*** The main advantages proposed by spin waves for data processing are briefly mentioned in the beginning and they were discussed in more details throughout the entire chapter of the book. Among others one can underline [19]: (*i*) efficient wave-based computing, (*ii*) small loss in insulating materials, (*iii*) wide frequency range up to THz, (*iv*) fundamental size limitations are given by the lattice constant of a magnetic material, (*v*) pronounced nonlinear phenomena, (*vi*) possibility of wire-less spin-wave excitation and detection, and (*vii*) rich spin-wave physics toolbox opening new ways for the transport and processing of data.

***Typical sizes and operational frequencies of magnonic devices.*** Nowadays, spin waves are studied experimentally in structures from millimeter down to micrometer sizes [4, 17, 19, 33, 54, 72]. The smallest reported prototypes reported so far have lateral sizes of few hundreds nanometers [42, 210, 224, 225]. At the same time, spin-wave devices studied by means of numerical simulation demonstrate promising functionalities at lateral sizes scaled down to at least tens of nanometers [26, 159, 172, 189]. The frequencies of the spin waves are mainly defined by the choice of the magnetic material, by the applied magnetic field and by the wavelengths of the spin wave. Nowadays, spin waves are usually investigated within the frequency range from one GHz to one hundred GHz. Nevertheless, there is intensive work going on to further reduce the size of the devices down to a few tens of nanometers and to increase the operating frequencies to sub-THz and THz frequency ranges.

***The long-term perspectives of spin-wave data processing.*** The perspectives are defined by the potential parameters of future devices, which, consequently, depend on the spin-wave characteristics. The qualitative analysis of the characteristics of exchange spin waves of nanometer wavelengths is performed in the section on spin waves in thin-film waveguides. It was shown e.g. that a spin wave of 5 nm wavelength has a frequency of about 2 THz in YIG and can propagate a distance of more than 16 μm which is more than three thousand times the wavelength. Therefore, spin waves indeed appear to be a promising candidate for the use as an information carrier in future ultrafast low-loss computing systems.

***Spin-wave logic devices that have already been demonstrated.*** At least, the following types of spin-wave logic devices have been realized experimentally at the level of proof-of-concept prototypes: (*i*) spin-wave logic gates in which spin-waves are manipulated by DC electric currents [143] (hardly suitable for the realization of an integrated circuit), (*ii*) spin-wave majority gate [25] (suitable for the realization of integrated magnonic circuits after the development and implementation of an energy efficient nonlinear amplitude normalizer), and (*iii*) magnon transistor for all-magnon data processing [24] (the technology is self-consistent but requires, first of all, miniaturization to the nanometer scale), (*iv*) a technique for magnetic microstructure imaging [248], (*v*) a device for prime factorization using spin-wave interference [249], (*vi*) a micrometer-scale spin-wave interferometer suitable for logic operations [250]. Besides, there are many very promising theoretical investigations in this field that were discussed in the chapter.

***Energy consumption of spin-wave devices.*** Since the field of spin-wave computing is at an initial stage of its development, the energy consumption can only be estimated with a certain accuracy and this strongly depends on the concrete choice of computing approach. For example, the approach based on the magnon transistor (an estimated energy consumption for nanoscaled magnon transistor is 5 aJ per bit [24]) most likely will require more energy in comparison to a spin-wave majority gate [21, 25, 145] operating with linear waves. In the latter case, the amplitude of the spin-wave is limited from below by the level of thermal noise rather than by the thresholds of nonlinear processes. This, in combination with the usage of fast exchange waves in low-damping material such as YIG, should ensure the energy consumption on the aJ level. However, the majority gate approach still requires the realization of an amplitude normalizer in order to combine many gates into a circuit and, currently, it is hard to estimate the total energy consumption of a fully functioning device. The energy consumption of spin-wave devices based on STT or ME cells converters directly depends on the efficiency of the conversion between spin waves and electric signal. Nowadays, energy cost for injecting of information into magnonic elements dominate the energy loss within the magnonic system itself. As an example, in Ref. [27] the authors have estimated that 24 aJ will be needed to excite spin waves per switching of a 100 nm × 100 nm ME cell.

*Interfacing of spin-wave devices to CMOS.* Any new concept for data processing requires interfacing with existing electronics devices and this is a topic of the section on spin-wave excitation, amplification and detection. Even in the case of all-magnon computing discussed above (when all information is kept inside of the magnonic system) the data should be transformed from electric form into magnon-carried signals in the magnonic circuit at the beginning, and should be read out at the end. In the case of other approaches, in which data is coded to and from magnonic elements after each operation [23], the interfacing is much more crucial. In particular, the efficiency of the conversion will define the energy consumption of such spin-wave devices. Therefore, the realization of efficient convertors by different means represents one of the most challenging task of modern magnon spintronics.

*Main challenges in the field of magnonics.* This book chapter, of course, only covers a chosen selection of scientific, engineering and technological problems. Among them probably the most important challenges are the miniaturization of the magnonic devices down to 10 nm sizes, increase of operating frequency to sub-THz and THz frequency ranges, the development of the approaches for the excitation and the detection of short-wavelength exchange magnons, development of low-loss and non-reciprocal spin-wave conduits, reflectionless guiding of spin waves in two dimensions, realization of highly-efficient spin-wave splitters and combiners, development of highly-efficient means for spin-wave amplification, investigations of non-linear spin-wave phenomena at the nanoscale and, as it was discussed above, the development of efficient converters between spin waves and electric signals.

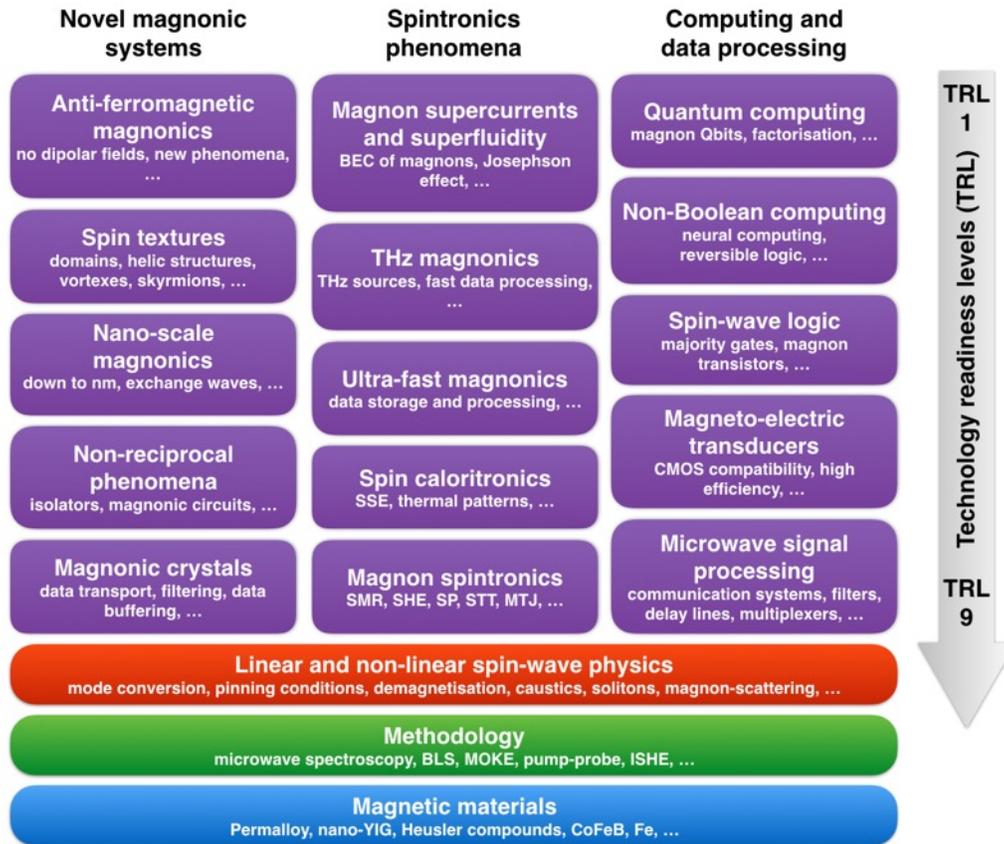

**Figure 12.** The variety of problems in modern magnonics and magnon spintronics.

*Research directions beyond the scope of this book chapter.* The fields of magnonics and magnon spintronics are very versatile and consist of many different research directions. A rough sketch of the selected research sub-fields is given in Figure 12. Some of the directions shown are already well established, other are just at the initial stage of their development but demonstrate much potential. In addition, one can see in the figure that the material science, the development of the methodology as well as the investigations of the physical spin-wave phenomena form the basis for all the research directions.

## Acknowledgments

I am grateful to Thomas Brächer, Burkard Hillebrands, Marjorie Lägel, Philipp Pirro, Oleksandr Serha, and Qi Wang for the inspiring discussions and for the support in the preparation of this chapter. The financial support by the ERC Starting Grant 678309 MagnonCircuits and DFG within Spin+X SFB/TRR 173 is strongly acknowledged.